\documentclass{article}
\usepackage[english]{babel}
\usepackage{amsmath,amssymb}
\usepackage{comment}
\usepackage{amsfonts} 
\usepackage[margin=1in]{geometry}
\usepackage{graphicx} 
\usepackage[colorlinks=true, allcolors=blue]{hyperref}
\usepackage{amsthm}
\usepackage{thm-restate}
\usepackage{tabularx}
\usepackage{enumitem}
\usepackage{booktabs}
\usepackage{caption}
\usepackage{framed}
\usepackage{ltablex}
\usepackage{ltxtable}
\usepackage{longtable}
\usepackage{microtype}
\usepackage{mleftright}
\usepackage{multirow}
\usepackage{natbib}
\usepackage{xspace}
\usepackage{xcolor}
\usepackage{nicefrac}
\usepackage{graphicx}
\usepackage{ifthen}
\usepackage{pifont}
\usepackage{pgfplots}
\newtheorem{theorem}{Theorem}

\newtheorem{proposition}{Proposition}

\newtheorem{lemma}{Lemma}
\newtheorem{example}{Example}

\newtheorem{claim}{Claim}

\newtheorem{definition}{Definition}

\usepackage{tikz}
\usetikzlibrary{positioning,arrows.meta,fit,backgrounds,decorations.pathreplacing,calc,matrix}

\usepackage{natbib}
\usepackage{algorithm}
\usepackage{algorithmic}

\newcommand{\pjrp}{PJR+\xspace}

\newcommand{\rpjrp}{Rank-\pjrp}
\newcommand{\psc}{PSC\xspace}

\newcommand{\psvc}{PSVC\xspace}
\newcommand{\pmvc}{PMVC\xspace}
\newcommand{\spsc}{Strong-PSC\xspace}
\DeclareMathOperator{\rank}{rank}

\newcommand{\rev}{\mathit{rev}}

\newcommand{\Seqf}{f^\mathit{seq}}
\newcommand{\tp}{\mathop{\mathtt{top}}}

\usepackage[capitalise,nameinlink]{cleveref}
\crefname{claim}{Claim}{Claims}
\crefname{observation}{Observation}{Observations}

\AddToHook{cmd/appendix/before}{%
  \crefalias{section}{appendix}%
}

\title{Veto Proportionality in Multiwinner Voting with Ranked Preferences}
\author{Haris Aziz\footnote{UNSW Sydney. Email: \texttt{haris.aziz@unsw.edu.au}}, Yuhang Guo\footnote{UNSW Sydney. Email: \texttt{yuhang.guo2@unsw.edu.au}}, and Patrick Lederer\footnote{ILLC, University of Amsterdam. Email: \texttt{p.lederer@uva.nl}}}
\date{}

\newtheoremstyle{bfnote}
{}{}%
{}{}%
{\bfseries}{.}%
{ }%
{\thmname{#1}\thmnumber{ #2}\thmnote{ (#3)}}
\theoremstyle{bfnote}
\newtheorem{remark}{Remark}

\begin{document}
	\maketitle
	\begin{abstract}
		A central issue in committee voting with ranked preferences is proportional representation: each group of voters should exert an influence on the selected committee that is proportional to its size. In the literature, this idea is formulated by a range of axioms, such as proportionality for solid coalitions (\psc), which demand that like-minded groups of voters are entitled to elect a fair number of candidates to the committee. However, in applications where voters are collectively affected by the entire committee, it may be equally important that no group is confronted with a disproportionate number of candidates that it regards as undesirable. Motivated by such applications, we initiate the study of veto proportionality in committee elections with ranked preferences. Our central notion, the proportional multiwinner veto core (\pmvc), generalizes Moulin's proportional singlewinner veto core to multiwinner elections by allowing groups of voters to partially veto sets of candidates. As our main contribution, we analyze how the \pmvc relates to existing fairness notions and design rules that are guaranteed to select committees from the \pmvc. Specifically, we first introduce a duality framework that connects inclusion- and exclusion-based fairness notions and show, e.g., that the dual axiom of the \pmvc strengthens \psc. Furthermore, we show that STV satisfies the dual axiom of the \pmvc, which also implies the existence of committees in the \pmvc. Moreover, subject to a strengthening of committee monotonicity, we characterize the rules that always return committees in the \pmvc as sequential adaptations of singlewinner voting rules that always select from the proportional singlewinner veto core. By our duality framework, this also gives rise to a new class of rules that satisfy \psc. We further prove that the dual axiom of the \pmvc and \rpjrp are jointly satisfiable, thereby pushing the frontier of fairness guarantees that can always be attained. Finally, we study the computational complexity of the \pmvc and show that deciding whether a given committee belongs to the \pmvc is coNP-complete.
	\end{abstract}
	
	\section{Introduction}
	
	In multiwinner elections, the goal is to select a fixed-size set of candidates based on the voters' preferences over the candidates. This problem has attracted significant attention in the last 10 years \citep[see, e.g.,][]{FSST17a,LacknerSkowrongABCvotingBook2023}, because it underlies a diverse range of applications, such as the election of parliaments or city councils \citep{ChCo83a}, participatory budgeting \citep{RSM25a}, recommender systems \citep{SFL16a}, and multi-criteria decision making \citep{boehmer2023subset}. One of the central considerations for all of these applications is proportional representation: every group of like-minded voters should exert an influence on the chosen committee that is proportional to the size of the group. This is maybe most intuitive for parliamentary elections, where, for example, a group comprising 20\% of the electorate should be able to secure roughly 20\% of the seats.
	
	In the literature on committee voting with ranked preferences, such notions of proportional representation have been formalized with a number of axioms \citep[e.g.,][]{Dumm84a,CJMW20a,azizlee2021,BrPe23a}, the most prominent of which is Dummett's Proportionality for Solid Coalitions (PSC). Intuitively, this axiom demands that, for every sufficiently large group of voters that ranks a set of candidates above all remaining candidates, an appropriate number of candidates from this set should be elected. Likewise, all other existing notions are also inclusion-focused: they demand that each group of voters is represented by a proportional number of candidates that they rank highly. This is particularly compelling when the committee members are viewed as representatives of the voters (e.g., in parliamentary elections) or as possible choices from a menu. A simple example of the latter is organizing a buffet with multiple dishes based on the participants' preferences. In this situation, it is desirable that the selected dishes proportionally represent the voters' preferences so that every participant finds food she likes, and it does not matter if the selected committee contains dishes that some participants dislike as such dishes can just be ignored. 
	
	However, there are numerous applications of committee voting in which voters are affected by the selected committee as a whole, rather than represented by individual committee members. For example, suppose that a department wants to adopt ten binding guidelines for the supervision of its PhD students. Different members of the department may evaluate the proposed guidelines differently, and it seems appropriate to apply the principles of proportional representation so that the resulting code fairly reflects the diverse perspectives. Here, however, proportional representation has a fundamentally different meaning: it is not enough that each group supports some of the selected guidelines. Instead, because every member must abide by the entire code, it may be even more important that no group strongly objects to a disproportionate number of them. Hence, proportionality refers here to a veto-based concept: each group of voters should be guaranteed that not too many of its disliked candidates are included. Moreover, we see numerous applications where such exclusion-focused proportionality ideas seem compelling.
	\begin{itemize}
		\item When the alternatives are public bads, inclusion-based fairness notions seem inappropriate. For example, say that a board needs to decide on funding cuts to several programs. In this case, it seems awkward to represent a voter by including their favorite proposals, as these are merely the funding cuts that are least painful to the voter. Instead, it seems much more important that no group of board members is disproportionally disadvantaged by the funding cuts. Similar concerns are, e.g., studied in obnoxious facility location \citep[e.g.,][]{ChurchDrezner2022,LamEtAl2024} and in the literature on harmful decisions \citep[e.g.,][]{CooksonEtAl2026}.
		\item Similar to our initial example of joint guidelines, we believe that exclusion-based proportionality is useful whenever voters need to agree on a joint statement that summarizes a variety of different points. Indeed, in such situations, it seems more important not to include positions that some voters strongly oppose than to give each voter some statements they favor highly. Moreover, if the participants are strongly polarized, it seems plausible that inclusion-based proportionality notions may even force us to include contradictory points in the joint statement. In addition to traditional settings, such joint statements are increasingly generated automatically in processes built around tools like Remesh \citep{KonyaEtAl2023}, which have recently started to gain attention in the computational social choice community \citep{FishEtAl2026,ChooiEtAl2026}. In particular, \citet{ChooiEtAl2026} formalize what it means for a voting rule to find common ground among LLM-generated statements via the proportional (singlewinner) veto core, and we will generalize this approach to multiwinner elections.
		\item As a third example, one may consider the creation of shared itineraries and playlists. In such applications, every participant has to take part in every activity on the itinerary or listen to all the songs on the playlist, so it seems reasonable to ensure that no group of voters dislikes a disproportionate amount of the activities or songs. For instance, this idea is reflected in the literature on group recommender systems through least-misery strategies, which aim to avoid recommendations that some participants strongly dislike \citep[e.g.,][]{Masthoff2004,ChaoEtAl2005,AgarwalEtAl2017}. In particular, while it may also be desirable that every group particularly enjoys some of the selected items, this alone does not prevent a group from disliking the overall itinerary or playlist for having too many undesirable items. 
	\end{itemize}

	Motivated by such applications, in this paper, we initiate the study of veto proportionality axioms for committee elections with ranked preferences. At a high level, veto proportionality requires that sufficiently large groups of voters can exclude candidates from the selected committee that they collectively regard as undesirable. To formally define this concept, we generalize Moulin's proportional veto core \citep{Mou81a,Moul82a} from singlewinner elections to multiwinner elections. Intuitively, the proportional singlewinner veto core assigns each group of voters a veto power proportional to its size and requires that no group can use this veto power to eliminate every candidate except candidates that all members of the group prefer to the chosen candidate. Or, put differently, no group should be able to use its veto power to guarantee the selection of a candidate that every member of the group prefers to the current winner. Our central veto proportionality notion, the \emph{proportional multiwinner veto core (\pmvc)}, extends this idea to committee elections by allowing groups to partially veto sets of candidates. In more detail, a group of voters can demand that at least $\ell$ candidates from a set $X$ be excluded from the committee if, by exerting its veto power, it can ensure that some other candidates that are unanimously preferred to $X$ by the members of the group are chosen instead. For example, with five candidates and two seats, a strict majority that ranks two candidates $y_1$ and $y_2$ above two other candidates $x_1$ and $x_2$ can use its proportional veto power to eliminate the fifth candidate $z$ and one of $x_1$ and $x_2$. Hence, the \pmvc requires for this profile that at least one of $x_1$ and $x_2$ is excluded from the chosen committee. Since the proportional singlewinner veto core has recently attracted considerable attention \citep[e.g.,][]{KoIa24a,KiKe23a,KiKe25a,HPS25a,ChooiEtAl2026}, we believe that its extension to multiwinner elections is also of independent interest.
	
	\subsection{Our Contributions}
	
	In this paper, we study the concept of veto proportionality through the lens of the proportional multiwinner veto core. Specifically, we first analyze how the proportional multiwinner veto core, and, more generally, veto proportionality, relate to existing fairness notions. Next, we derive several insights into how committees in the proportional multiwinner veto core can be computed and how this notion interacts with other desirable properties. Lastly, we turn to computational questions concerning the verification of whether a committee belongs to the \pmvc. In more detail, we show the following results (see also \Cref{fig:landscape}).
	
	\begin{enumerate}[label=(\arabic*), leftmargin=*]
		\item First, to relate the proportional multiwinner veto core to existing fairness notions, we introduce a duality framework that allows us to map inclusion-based proportionality axioms to exclusion-based ones and vice versa. The idea underlying this framework is that the candidates that an inclusion-based proportionality notion requires us to choose for a profile are precisely those that should be excluded when reversing the preference profile. More formally, a committee $W$ satisfies the dual of a fairness notion for a profile~$R$ and a committee size~$k$ exactly if the committee $C\setminus W$ satisfies the original fairness notion for the reversed profile $\rev(R)$ and the committee size $m-k$. We then show that the dual axiom of the proportional multiwinner veto core, which we call \spsc, implies proportionality for solid coalitions, is incomparable with \rpjrp by \citet{BrPe23a}, and is implied by a variant of core-stability (\Cref{prop:landscape}), thereby relating our new notion to established inclusion-based proportionality axioms. 
		
		\item Next, we analyze how to compute committees in the proportional multiwinner veto core. To this end, we show that a classical rule called STV satisfies \spsc, whereas other known rules, such as the Expanding Approvals Rule by \citet{AzLe19a} and the Solid Coalition Refinement Rule by \citet{ALP+25a}, fail this condition (\Cref{prop:classicrules}). This means that the dual rule of STV returns committees in the \pmvc, thereby showing that our central concept is always satisfiable. Moreover, the fact that STV satisfies \spsc while the other rules fail this condition gives an argument for STV over these rules.
		
		\item We prove that singlewinner rules that always return candidates in the proportional singlewinner veto core can be used to find committees in the \pmvc by repeatedly adding their selected candidate to the committee and deleting it from the profile (\Cref{thm:general_seq_single_winner_veto_pmvc}). Moreover, subject to a strengthening of committee monotonicity that we call committee expansion consistency, we show that these are even the only rules that always compute committees in the \pmvc (\Cref{thm:characterization_mpvc_consistency}). We note that, by our duality framework, this also means that we obtain a large class of rules that satisfy \spsc and thus \psc, thereby providing a much clearer picture of how these axioms can be satisfied. In particular, it is known that a candidate belongs to the singlewinner veto core if and only if it can be selected by a Serial Veto rule \citep{Moul83a,KiKe23a}, so we obtain a functional form characterization of these rules. Using this description, we also single out the Simultaneous Veto rule as the Serial Veto rule induced by a particularly fair veto sequence (\Cref{thm:simultaneousVeto}).
		
		\item As our technically most challenging result, we consider whether the logically incomparable notions of \spsc and \rpjrp (or, equivalently, \pmvc and the dual of \rpjrp) can be jointly satisfied. Using a repair scheme based on a novel budgeting approach, we show that committees that simultaneously satisfy both properties always exist (\Cref{thm:spsc_rpjrp_compatibility}). This proves that two of the most demanding fairness notions that are individually always satisfiable can in fact be satisfied simultaneously, thereby pushing the frontier of fairness guarantees that can be attained in every election.
		
		\item Lastly, we study the computational complexity of deciding whether a given committee satisfies \pmvc or \spsc for a profile. This problem is relevant, for example, when auditing whether a chosen committee is fair in practice or when experimentally analyzing how often a voting rule chooses committees satisfying these notions. Unfortunately, it turns out that this verification task is coNP-complete (\Cref{thm:verification}, which is shown by reduction from vertex cover on cubic graphs. On the positive side, we prove that whether these axioms are violated with respect to a particular set of candidates or a given set of voters can be decided in polynomial time (\Cref{prop:FPT}), which allows us to construct FPT algorithms parameterized by the number of candidates and voters, respectively.
	\end{enumerate}
	
	\begin{figure}[t]
		\centering
		\tikzset{
			vp box/.style={draw=black!80, line width=0.4pt, align=center, inner sep=2.5pt, font=\scriptsize,
				execute at begin node={\hyphenpenalty=10000\exhyphenpenalty=10000}},
			vp new/.style={fill=black!8},
			vp imp/.style={-{Stealth[length=1.9mm,width=1.5mm]}, line width=0.5pt},
			vp inc/.style={dashed, line width=0.5pt, black!60},
		}
		\newcommand{\vpName}[1]{{\footnotesize\bfseries #1}\\[1pt]}
		\begin{tikzpicture}[x=1cm, y=1cm]
			\node[vp box, text width=26mm] (core) at (4.2,5.35) {\vpName{Core-stability}\itshape may be empty};
			\node[vp box, text width=41.5mm] (jnt) at (4.2,3.95) {\vpName{\spsc{} + \rpjrp}(Existence)~\Cref{thm:spsc_rpjrp_compatibility}};
			\node[vp box, vp new, text width=36.5mm, anchor=north] (sp) at (1.913,2.85) {\vpName{\spsc}STV\\ rev.-seq.\ Ordered Phragm\'en
				\\ dual seq.\ Serial Veto
			};
			\node[vp box, text width=26.5mm, anchor=north] (rp) at (6.489,2.85) {\vpName{\rpjrp}EAR};
			\node[vp box, text width=26mm] (psc) at (4.2,0.5) {\vpName{\psc}SCR
			};
			\draw[vp imp] (core) -- (jnt);
			\draw[vp imp] (jnt) -- (sp);
			\draw[vp imp] (jnt) -- (rp);
			\draw[vp imp] (sp) -- (psc);
			\draw[vp imp] (rp) -- (psc);
			\draw[vp inc] (sp.north east) -- node[above=0.5pt, font=\scriptsize\itshape, text=black!70] {incomparable} (rp.north west);
			\node[vp box, text width=26mm] (dcore) at (12.103,5.35) {\vpName{Dual core}\itshape may be empty};
			\node[vp box, text width=41.5mm] (djnt) at (12.103,3.95) {\vpName{\pmvc{} + dual \rpjrp}(Existence)~\Cref{thm:spsc_rpjrp_compatibility}};
			\node[vp box, text width=26.5mm, anchor=north] (drp) at (9.815,2.85) {\vpName{Dual \rpjrp}dual EAR};
			\node[vp box, vp new, text width=36.5mm, anchor=north] (pm) at (14.391,2.85) {\vpName{\pmvc}dual STV\\ seq.\ Simultaneous Veto
				\\ seq.\ Serial Veto
			};
			\node[vp box, text width=26mm] (dpsc) at (12.103,0.5) {\vpName{Dual \psc}dual SCR
			};
			\draw[vp imp] (dcore) -- (djnt);
			\draw[vp imp] (djnt) -- (drp);
			\draw[vp imp] (djnt) -- (pm);
			\draw[vp imp] (drp) -- (dpsc);
			\draw[vp imp] (pm) -- (dpsc);
			\draw[vp inc] (drp.north east) -- node[above=0.5pt, font=\scriptsize\itshape, text=black!70] {incomparable} (pm.north west);
			\node[font=\small\bfseries, anchor=base] (hl) at (4.2,6.15) {Inclusion-based Proportionality};
			\node[font=\small\bfseries, anchor=base] (hr) at (12.103,6.15) {Exclusion-based Proportionality};
			\draw[{Stealth[length=1.9mm,width=1.5mm]}-{Stealth[length=1.9mm,width=1.5mm]}, line width=0.5pt]
			([xshift=4pt]hl.east) -- node[above=1pt, font=\footnotesize\itshape] {duality} ([xshift=-4pt]hr.west);
			\draw[black!30, line width=0.4pt] (8.152,5.85) -- (8.152,0.0);
		\end{tikzpicture}
		\caption[Landscape of proportionality axioms and the rules satisfying them]{Summary of inclusion-based (left) and exclusion-based (right) proportionality axioms. Arrows denote implications (\Cref{prop:landscape}; \citealp{BrPe23a}); duality (\Cref{def:dual_axiom}) maps each axiom and rule to its mirror image on the other side and preserves implications and rule guarantees (\Cref{prop:dualimplication}). The shaded box marks our conceptual contribution. Each rule is listed at the strongest axiom it is known to satisfy (\Cref{prop:classicrules,thm:general_seq_single_winner_veto_pmvc}; \citealp{BrPe23a,ALP+25a}). 
		Seq.\ denotes the sequential extension; reverse-sequential (rev.-seq.) Ordered Phragm\'en \citep{ALP+25a} is the dual of seq.\ Simultaneous Veto.}
		\label{fig:landscape}
	\end{figure}

	\subsection{Related Work}
	Our work is primarily related to two strands of literature: the study of proportional representation in multiwinner voting with ranked preferences and of proportional veto power in social choice theory. We next provide an overview of these two strands. 
	
	\paragraph{Proportional Representation in Committee Voting.}
	Proportional representation is a central objective in committee voting. For ranked preferences, this idea was first formalized by Dummett's {proportionality for solid coalitions} (\psc), which guarantees that so-called solid coalitions are fairly represented in the selected committee \citep{Dumm84a,Wood94a}. Several well-known rules satisfy \psc, including the Single Transferable Vote (STV), the Quota Borda System of \citet{Dumm84a}, and the Expanding Approvals Rule (EAR) of \citet{AzLe19a}. \citet{AzLe19a,azizlee2021} extended \psc to settings with weak preferences and participatory budgeting through notions such as generalized PSC and inclusion PSC. Moreover, \citet{AzLe22a} characterized the committees satisfying \psc as precisely those that can be returned by a class of Minimal Demand rules. A closely related and substantially larger literature studies proportional representation under approval preferences \citep[e.g.,][]{ABC17a,Fernandez_Elkind_Lackner_Fernandez_Fisteus_Basanta_Val_Skowron_2017,methodofequalshares}. This literature has introduced axioms such as justified representation (JR) and proportional justified representation (PJR), together with numerous refinements and strengthenings \citep[see][for an overview]{LacknerSkowrongABCvotingBook2023}. Recently, \citet{BrPe23a} adapted these approval-based fairness notions to committee voting under ranked preferences by requiring that the chosen committee satisfy the corresponding axioms in the approval profiles induced by all rank thresholds. Moreover, they showed that one of these notions, called \rpjrp, strengthens \psc and is satisfied by EAR. Furthermore, \citet{ALP+25a} studied the compatibility between \psc and committee monotonicity. In particular, they introduced the Solid Coalition Refinement (SCR) rule and reverse sequential rules and showed that these rules satisfy both desiderata. Complementing this axiomatic literature, \citet{BardalEtAl2025} introduced quantitative measures of proportionality for ranked elections and used them to compare voting rules on real-world election data. Another related approach formalizes proportional representation through core-stability, which requires that no group of voters can deviate to a committee of proportional size that all members of the group prefer. Since such committees need not always exist, prior work has also investigated approximate variants of this requirement~\citep{CJMW20a,JMW20a}. In contrast to our work, all of these papers focus on inclusion-based fairness notions, whereas our primary concern is exclusion-based proportionality.
	
	\paragraph{Proportional Veto Power.}
	The idea that groups of voters should possess veto power in proportion to their size originates from the proportional veto principle of \citet{Mou81a}. In the singlewinner setting with ranked preferences, Moulin formalized this idea through the proportional veto core, established that this core is always nonempty, and developed procedures for selecting candidates from it \citep{Mou81a,Moul82a}. More recently, \citet{KoNe20a} compared the majority and veto power of different voting rules, while \citet{IaKo21a} showed that the proportional veto core can be computed in polynomial time. Subsequent work has studied the axiomatic properties of rules that select from the veto core \citep{KoIa24a} and connections between veto-core constructions, voting rules, and metric distortion \citep{Pete23a,KiKe23a,KiKe25a}. In particular, \citet{Pete23a} observed a connection between the proportional veto core and proportionality for solid coalitions under preference reversal and committee complementation in the special case of committees containing all but one candidate. Proportional veto ideas have also been adapted to continuous alternative spaces, with applications to federated learning, policy aggregation, and the generation of common-ground statements \citep{RMY+24a,AEP24a,ChooiEtAl2026}. A related literature incorporates negative preferences directly into multiwinner voting: \citet{TaPa21a} and \citet{KPS+25a} study proportional committee selection under trichotomous preferences and investigate proportional veto guarantees separately from positive representation guarantees. Moreover, \citet{HPS25a} extend Moulin's proportional veto principle to approval ballots and multiwinner elections through flexible voter representation. Our work complements these developments by studying proportional veto power for committee elections with ranked preferences and by relating proportional exclusion systematically to classical inclusion-based notions of proportional representation.
	
	\section{The Model}
	
	In this section, we first discuss the basic framework of multiwinner elections and then move on to our central veto proportionality notion, the proportional multiwinner veto core. 
	
	\subsection{Multiwinner Elections}
	Throughout the paper, we consider the standard multiwinner voting model with strict preferences. Let $N=\{1,\ldots, n\}$ denote a set of $n$ voters and let $C=\{c_1,\ldots, c_m\}$ denote a set of $m$ candidates. Each voter $i\in N$ is assumed to have a \emph{strict preference relation $\succ_i$}, which is formally a complete, transitive, and asymmetric binary relation over all candidates in $C$. A \emph{(preference) profile} $R={(\succ_i)}_{i\in N}$ is the collection of the voters' preference relations. We typically write preference relations as comma-separated lists and indicate for preference profiles only how many voters report a given preference relation. For example, $3: x_1,x_2,x_3$ means that $3$ voters prefer candidates $x_1$ to $x_2$ to $x_3$. Furthermore, given a set of voters $S\subseteq N$ and two sets of candidates $X,Y\subseteq C$, we write $X\succ_S Y$ to mean that all voters in $S$ prefer all candidates in $X$ to all candidates in $Y$. That is, $X\succ_S Y$ if and only if $x\succ_i y$ for all $i\in S$, $x\in X$, and $y\in Y$. Moreover, if $X$ or $Y$ is a singleton set, we omit the set brackets and only write, e.g., $x\succ_S Y$ instead of $\{x\}\succ_S Y$.
	
	Given a preference profile, our goal is to select a \emph{committee}, which is formally a subset of the candidates of a given size $k$. To this end, we will use \emph{committee voting rules}, denoted by $f$, which map every profile $R$ and target committee size $k$ to a single winning committee $f(R,k)=W\subseteq C$ with $|W|=k$. In particular, we define $f(R,0)=\emptyset$ and $f(R,m)=C$. 
	
	\subsection{The Proportional Multiwinner Veto Core} 
	We now introduce the central axiom of this paper, the proportional multiwinner veto core. Since this axiom is inspired by the proportional singlewinner veto core by \citet{Mou81a}, we first define this concept. Intuitively, the proportional singlewinner veto core guarantees each voter a veto power of marginally less than $\frac{m}{n}$. Then, this axiom requires that we do not choose a candidate such that all voters in some group can get better off by redistributing their veto power. In more detail, for a candidate $x$ to be in the proportional singlewinner veto core, there should not be a set of voters $S$ and a set of candidates $Y$ such that all voters in $S$ prefer all candidates in $Y$ to $x$ and the set $S$ is large enough to veto all candidates outside of $Y$. Indeed, if $x$ were chosen, the group $S$ would be better off by reallocating their veto power to veto all candidates outside of $Y$.
	
	\begin{definition}[Proportional Singlewinner Veto Core (\psvc)~\citep{Mou81a}]
		\label{def:vetocore}
		A candidate $x$ is \emph{blocked} by a group of voters $S\subseteq N$ if there exists a set of candidates $Y\subseteq C$ such that $|S|>\frac{n}{m}\cdot |C\setminus Y|$ and $Y\succ_S x$. The \emph{proportional (singlewinner) veto core} is the set of candidates that are not blocked by any group of voters.
	\end{definition}
	
	To extend this axiom to multiwinner elections, we will make two changes. First, since we need to choose $k$ candidates instead of one, the veto power of each voter will be (marginally less than) $\frac{m-k+1}{n}$ instead of $\frac{m}{n}$. This is necessary to ensure that there is always a committee satisfying the resulting notion. Second, we will allow for voters to partially veto sets of candidates. In more detail, it should not be possible that a group of voters $S$ can, by redistributing its veto power, replace some candidates in the considered committee $W$ with candidates that every voter of the group finds more desirable. Formally, we capture this with the notion of $\ell$-blocked sets, discussed in the following definition.
	
	\begin{definition}[Proportional Multiwinner Veto Core (\pmvc)]
		\label{def:pmvc}
		A set of candidates $X$ is \emph{$\ell$-blocked} by a non-empty group of voters $S\subseteq N$ if there exists a set of candidates $Y\subseteq C$ such that $|S|> \frac{n}{m-k+1}(|C\setminus Y|-(|X|-\ell))$ and $Y\succ_S X$. A committee $W$ is in the proportional multiwinner veto core (\pmvc) for a profile $R$ and target committee size $k$ if $|X\setminus W|\geq \ell$ for every set of candidates $X$ and integer $\ell\in\{1,\dots, |X|\}$ such that $X$ is $\ell$-blocked by some group of voters. 
	\end{definition}
	
	Put differently, this axiom demands that if the group of voters $S$ is large enough to veto all but $(|X|-\ell)$ candidates from $C\setminus Y$, then at least $\ell$ candidates from $X$ are not in the chosen committee. Indeed, the group $S$ would veto any candidate in $X$ before those in $Y$ since every voter in $S$ prefers all candidates in $Y$ to all candidates in $X$. Hence, by the pigeonhole principle, at least $\ell$ candidates from $X$ must be vetoed by these voters. We further note that, for $k=1$, the proportional multiwinner veto core reduces to the proportional singlewinner veto core. Finally, the proportional multiwinner veto core can also be defined without the notion of $\ell$-blocked sets of candidates. In more detail, an equivalent definition is the following: a committee $W$ is in the \pmvc for a profile $R$ and a target committee size $k$ if $|W\cap X|\leq \max(0, m-\ell-|Y|)$ for all sets of voters $S$, integers $\ell$, and sets of candidates $X$ and $Y$ such that $|S|>\frac{n}{m-k+1}\ell$ and $Y\succ_S X$.
	
	We will next consider an example illustrating the proportional multiwinner veto core.
	\begin{example}
		\label{example:pmvc}
		Consider the following profile with $6$ voters and $6$ candidates for the committee size $k=3$.
		
		\begin{center}
			\begin{tabular}{cc}
				1: & \, a, b, c, d, e, f \\
				1: & \, b, a, f, d, e, c \\
				1: & \, f, a, b, c, d, e \\
				1: & \, b, f, a, c, e, d \\
				1: & \, a, b, e, d, c, f \\
				1: & \, e, d, c, f, b, a 
			\end{tabular}
		\end{center}
		We will next show that the committee $W=\{a,d,e\}$ is not in the \pmvc. To see this, let $X=\{c,d,e\}$ and $Y=\{a,b\}$. We observe that $X$ is $2$-blocked by the group of the first $5$ voters as everyone ranks $Y$ over $X$ and $5 > \frac{6}{6-3+1}(4 - (3-2))=\frac{9}{2}$. \pmvc thus requires that at least $2$ candidates from $X$ are excluded from the winning committee. However, both $d$ and $e$ are elected by $W$, which implies that this committee is not in the proportional multiwinner veto core. By contrast, the committee $\{a,b,d\}$ is in the proportional multiwinner veto core as no such blocked subset of candidates exists. In fact, there are $3$ committees that are in \pmvc, which are $\{a,b,c\}$, $\{a,b,d\}$, and $\{a,b,e\}$ as $f$ is $1$-blocked by the $2$ voters who rank it last. 
	\end{example}
	
	\section{From Proportional Representation to Veto Proportionality}
	
	In this section, we relate the proportional multiwinner veto core to existing fairness notions such as proportionality for solid coalitions (PSC) and core-stability. To this end, in \Cref{subsec:duality} we introduce a duality framework that allows us to map inclusion-based proportionality axioms to veto-based proportionality axioms and vice versa. In \Cref{subsec:landscape}, we then show that the dual axiom of the \pmvc implies PSC, is implied by a variant of core-stability, and is logically incomparable with another fairness notion called \rpjrp.
	
	\subsection{The Duality Framework}\label{subsec:duality}
	
	To relate the \pmvc to existing notions of proportional representation, we next introduce our duality framework. For this, we roughly follow the setup of \citet{DoPe26a} and treat fairness notions as functions $F$ that map each profile $R$ and target committee size $k$ to a set of size-$k$ committees. Intuitively, $F(R,k)$ is the set of committees that satisfy the fairness notion for the given profile $R$ and committee size $k$. Or, put differently, in this section, we treat a fairness notion as the set-valued committee voting rule that returns all committees satisfying the given axiom.
	
	The idea of our duality framework is the following: the candidates that an inclusion-based fairness notion requires to be selected in a profile are exactly those that should not be selected under the corresponding veto-based notion in the reversed profile. Thus, if a fairness notion $F$ deems $W$ acceptable for a profile $R$, its dual should deem $C\setminus W$ acceptable for the reversed profile $\rev(R)$. The same construction can also be used to turn a veto proportionality notion, such as the \pmvc, into an inclusion-based notion. To formalize this idea, given a preference relation $\succ$, let $\rev(\succ)$ denote its reverse preference relation: for any pair of distinct candidates $x,y\in C$, it holds that $x\,\rev(\succ)\,y$ if and only if $y\succ x$. Analogously, for every preference profile $R=(\succ_i)_{i\in N}$, we denote by $\rev(R)=(\rev(\succ_i))_{i\in N}$ its reverse preference profile. The dual fairness notion of $F$ is then obtained by reversing the input profile, asking $F$ to select $m-k$ candidates for the reversed profile, and returning the $k$ unselected candidates.

	\begin{definition}[Dual Fairness Notion]
		\label{def:dual_axiom}
		Let $F$ be a fairness notion. The dual fairness notion of $F$, denoted by $F^d$, is defined by $W\in F^d(R,k)$ if and only if $C\setminus W\in F(\rev(R), m-k)$ for all committees $W$, profiles $R$, and target committee sizes $k$.
	\end{definition}
	
	To illustrate this definition, we consider again \Cref{example:pmvc}, focusing on the dual of the proportional multiwinner veto core for $k=3$. By the definition of duality, to find the committees satisfying $\pmvc^d$, we first reverse the preference profile and find the size-$(m-k)$ committees satisfying \pmvc in $\rev(R)$. An immediate observation is that, in $\rev(R)$, the first five voters all prefer $\{c,d,e\}$ to $\{a,b\}$, which means that $\{a,b\}$ is $2$-blocked. Hence, \pmvc requires that neither $a$ nor $b$ is elected. Therefore, for $k=3$, only size-$3$ subsets of $\{c,d,e,f\}$ can satisfy \pmvc in $\rev(R)$, and all such committees indeed satisfy this fairness condition. By taking the complements of these sets, we infer that the committees satisfying $\pmvc^d$ in $R$ are given by $\pmvc^d(R,3)=\{\{a,b,c\},\{a,b,d\},\{a,b,e\},\{a,b,f\}\}$.
	
	We conclude this section by giving some further insights into dual fairness notions. First, it is easy to see that the dual of the dual of a fairness axiom is the original fairness axiom. Indeed, it holds that $W\in (F^d)^d(R,k)$ if and only if $C\setminus W\in F^d(\rev(R), m-k)$ if and only if $W=C\setminus (C\setminus W)\in F(\rev(\rev(R)), m-(m-k))=F(R,k)$. Furthermore, we will next show that dual fairness notions satisfy the same logical relationships as the original axioms and that, if a rule satisfies a given fairness notion, its dual rule satisfies the corresponding dual fairness notion. To make this more precise, we first note that, in our set-valued notation, a fairness notion $F_1$ implies another fairness notion $F_2$ if $F_1(R,k)\subseteq F_2(R,k)$ for each profile $R$ and committee size $k$. That is, a fairness notion implies another one if every committee that satisfies the first notion also satisfies the second one. Further, we define the dual rule $f^d$ of a committee voting rule $f$ analogous to dual fairness notions: $f^d(R,k)=W$ if and only if $f(\rev(R), m-k)=C\setminus W$. This means we treat rules here essentially as single-valued fairness notions. We defer the simple proof of the subsequent proposition to Appendix~\ref{app:DualImplication}.
	
	\begin{restatable}{proposition}{DualImplication}\label{prop:dualimplication}
		The following claims hold:
		\begin{enumerate}[label=(\arabic*), leftmargin = *]
			\item For any two fairness notions $F$ and $G$, it holds that $G(R,k)\subseteq F(R,k)$ for all profiles $R$ and committee sizes $k$ if and only if $G^d(R,k)\subseteq F^d(R,k)$ for all profiles $R$ and committee sizes $k$.
			\item For every fairness notion $G$ and committee voting rule $f$, it holds that $f(R,k)\in G(R,k)$ for all profiles $R$ and committee sizes $k$ if and only if $f^d(R,k)\in G^d(R,k)$ for all profiles $R$ and committee sizes $k$.
		\end{enumerate}
	\end{restatable}
	
	\subsection{An Axiomatic Landscape}\label{subsec:landscape}
	
	Using the duality framework, we will now show that the proportional multiwinner veto core, or more precisely its dual, is related to existing fairness notions. Specifically, we will prove that the dual of the proportional multiwinner veto core implies the well-known notion of proportionality for solid coalitions \citep{Dumm84a,Tide95a}, is incomparable to \rpjrp~\citep{BrPe23a}, and is implied by a variant of core-stability \citep[e.g.,][]{ABC17a,CJMW20a,JMW20a}.
	
	To show these claims, we next derive the dual of the proportional multiwinner veto core. For this, we fix a profile $R$ and a committee size $k$. Now, \pmvc requires for the reverse profile $\rev(R)$ and the committee size $m-k$ that, for every $\ell$-blocked set of candidates $X$, at least $\ell$ candidates from $X$ are unchosen. For the original profile $R$, this means that at least $\ell$ candidates from $X$ are chosen. Furthermore, a set of candidates $X$ is $\ell$-blocked with respect to $\rev(R)$ and $m-k$ if there exists a set of voters $S$ and a set of candidates $Y$ such that $|S|>\frac{n}{m-(m-k)+1}(|C\setminus Y|-(|X|-\ell))$ and $Y \mathrel{\rev(\succ_S)} X$. For the original profile $R$, this means that there is a set of voters $S$, an integer $\ell\in \{1,\dots, |X|\}$, and a set of candidates $Y$ such that $|S|>\frac{n}{k+1}(|C\setminus Y|-(|X|-\ell))$ and $X\succ_S Y$. Hence, the dual of \pmvc requires of a committee $W$ that $|W\cap X|\geq \min(|X|, \ell)$ for all sets of voters $S$, sets of candidates $X,Y$, and integers $\ell$ such that $|S|>(\ell+|C\setminus (X\cup Y)|)\frac{n}{k+1}$ and $X\succ_S Y$. We call the resulting property strong proportionality for solid coalitions. 
	
	\begin{definition}[\spsc]
		A committee $W$ satisfies strong proportionality for solid coalitions (\spsc) for a profile $R$ and committee size $k$ if, for all sets of voters $S\subseteq N$, integers $\ell\in\mathbb{N}$, and sets of candidates $X,Y\subseteq C$ such that $|S|>(\ell+|C\setminus (X\cup Y)|)\frac{n}{k+1}$ and $X\succ_S Y$, it holds that $|W\cap X|\geq \min(|X|, \ell)$.
	\end{definition}
	
	Next, we introduce three established axioms of proportional representation, namely proportionality for solid coalitions, \rpjrp, and core-stability. We define all these axioms based on the Droop-quota $q=\frac{n}{k+1}$, whereas other works consider these notions under the Hare-quota $q=\frac{n}{k}$. The reason for this is that the proportional singlewinner veto core and thus also our proportional multiwinner veto core are defined based on the Droop-quota.

	\paragraph{Proportionality for Solid Coalitions.} Our first inclusion-based proportionality axiom is called proportionality for solid coalitions and goes back to \citet{Dumm84a}. The idea of the axiom is that, if there is a sufficiently large set of voters that all prefer a subset of candidates over the other candidates, then this group should be represented by a number of candidates from their preferred candidate subset that is proportional to the size of the group. To formalize this, we say a group of voters $S\subseteq N$ is a \emph{solid coalition} supporting a set of candidates $X\subseteq C$ if and only if $X\succ_S C\setminus X$. Then, a size-$k$ committee $W$ satisfies proportionality for solid coalitions (\psc) for a preference profile $R$ if $|W\cap X|\geq \min(\ell, |X|)$ for every set of candidates $X$ and integer $\ell$ such that $X$ is supported by a solid coalition $S$ of size $|S|> \ell \cdot \frac{n}{k+1}$.
	
	\paragraph{\rpjrp.} Next, we turn to \rpjrp, which is due to \citet{BrPe23a}. To introduce this notion, we define the rank of a candidate $c$ in a preference relation $\succ$ by $\rank(c,\succ)=1+|\{c'\in C\setminus \{c\}\colon c'\succ c\}|$. For example, the top-ranked candidate of a voter has rank $1$, her second top-ranked candidate has rank $2$, and so on. Then, \rpjrp requires that, if a group of voters is large enough to demand $\ell$ candidates, then for each rank $r$, it should either hold that the group does not have a common unselected candidate among their top-$r$ candidates or that at least $\ell$ of the top-$r$ candidates are selected. More formally, a size-$k$ committee $W$ satisfies \rpjrp for a profile $R$ if and only if there is no candidate $x \in C\setminus W$, rank $r \in [m]$, and group of voters $S$ such that $\lvert S \rvert > \frac{\ell n}{k+1}$ for some $\ell\in\mathbb{N}$, $\rank(x,\succ_i) \leq r$ for all $i \in S$, and $\lvert\{c \in W \colon \rank(c,\succ_i) \le r \text{ for some } i\in S\}\rvert < \ell$.
	
	\paragraph{Core-stability.} As our third condition, we turn to core-stability. Intuitively, this axiom demands that no group of voters can find a set of candidates whose size is proportional to that of the group such that each voter in our group prefers this set to the selected committee. To formalize this notion, we say a group of voters $S$ prefers a set $X$ to another set $Y$ if $X\setminus Y\neq \emptyset$, $Y\setminus X\neq \emptyset$, and $X\setminus Y\succ_S Y\setminus X$, i.e., all voters in $S$ prefer all candidates $X\setminus Y$ to those in $Y\setminus X$. Then, a size-$k$ committee $W$ is \emph{core-stable} for a profile $R$ if, for every group of voters $S\subseteq N$, there is no set of candidates $X$ such that $|S|>|X|\cdot \frac{n}{k+1}$ and all voters in $S$ prefer $X$ to $W$. We note that our definition of core-stability differs slightly from existing notions, since we use the Droop-quota and many other definitions merely require that voters compare two sets based on their favorite candidates in these sets. Despite these changes, examples from the literature \citep{ijcai2017p13,JMW20a} show that core-stable committees are not guaranteed to exist.\medskip
	
	Finally, we show the logical relations between our axioms; we refer to \Cref{fig:landscape} for a graphical summary of these relations. We note that the implication from \rpjrp to \psc was already noted by \citet{BrPe23a}. The other insights are, to the best of our knowledge, new. The proof of the following proposition can be found in Appendix~\ref{app:DualImplication}.
	
	\begin{restatable}{proposition}{landscape}\label{prop:landscape}
		The following claims hold: 
		\begin{enumerate}[label=(\arabic*), leftmargin=*]
			\item If a committee $W$ is core-stable for a profile, it also satisfies \rpjrp and \spsc.
			\item \spsc and \rpjrp are logically incomparable, in the sense that a committee satisfying \rpjrp may fail \spsc and a committee satisfying \spsc may fail \rpjrp. 
			\item If a committee $W$ satisfies \spsc or \rpjrp for a profile, it also satisfies \psc.
		\end{enumerate}
	\end{restatable}

	We note that \Cref{prop:landscape}, in combination with \Cref{prop:dualimplication}, yields two important insights. Firstly, by the logical implications derived in the last proposition, a natural attempt to satisfy \spsc is to consider rules that satisfy PSC. If such a rule would indeed satisfy \spsc, its dual would produce committees that belong to the proportional multiwinner veto core by Claim (2) of \Cref{prop:dualimplication}. Conversely, by studying rules that find committees in the proportional multiwinner veto core, we also get novel insights into PSC.
	
	\section{Satisfaction of the Proportional Multiwinner Veto Core}\label{sec:rules}
	
	We next turn to the question of how to find committees in the proportional multiwinner veto core and, more broadly, how this fairness axiom interacts with further requirements. Towards this, we will analyze four aspects. First, in \Cref{subsec:classicalRules}, we check for three established rules whether they satisfy \spsc, which answers whether their dual rules satisfy \pmvc. Second, in \Cref{subsec:newRules}, we adapt rules that find winners in the proportional singlewinner veto core to our setting and show that the resulting rules always return committees in the \pmvc. Third, based on insights from the singlewinner setting, we characterize in \Cref{subsec:characterization} a class of sequential committee voting rules that return committees in the proportional multiwinner veto core, thereby offering a more structural insight into the space of rules satisfying \pmvc. Finally, in \Cref{subsec:SpscAndRpjrp}, we show that, for each profile, there are committees that satisfy both \spsc and \rpjrp, which demonstrates that these demanding and logically incomparable fairness notions are jointly satisfiable. 
	
	\subsection{Classical Rules}
	\label{subsec:classicalRules}
	
	We will first check whether existing committee voting rules satisfy \spsc. By \Cref{prop:dualimplication}, if this is the case, the dual of these rules would be guaranteed to produce committees in the \pmvc. Specifically, we will consider three rules in this section: the Single Transferable Vote (STV) \citep{Dumm84a}, the Expanding Approvals Rule by \citet{AzLe19a}, and the Solid Coalition Refinement Rule (SCR) by \citet{ALP+25a}. We formally introduce these rules in the following. Since all of these rules may encounter ties between candidates during their execution, we assume that ties are broken by a fixed tie-breaking order. 
	
	\paragraph{Single Transferable Vote.} The \emph{Single Transferable Vote (STV)} is perhaps the most well-known voting rule for selecting proportional committees, as it was the first known rule to satisfy \psc \citep{Tide95a}. Given a profile $R$ and the target committee size $k$, it works as follows: each voter is initialized with a unit voting weight and in each round, we consider the candidate $x$ that is top-ranked by the voters with the largest total weight. If the total weight of these voters exceeds $\frac{n}{k+1}$, we add $x$ to the committee, remove it from the profile, and reduce the total weight of these voters by $\frac{n}{k+1}$ while keeping all weights nonnegative. If the total weight of voters who top-rank $x$ is at most $\frac{n}{k+1}$, we delete the candidate that is top-ranked by the voters with the smallest total weight from the profile, and the voters who top-ranked this candidate move on to their next-best candidate. This process is continued until $k$ candidates are chosen. We note that, to fully specify STV, one needs to define precisely which voters' weights are reduced by STV when a candidate is selected. However, our results will be independent of this aspect, so we do not go into further details here.
	
	\paragraph{Expanding Approvals Rule.} The \emph{Expanding Approvals Rule (EAR)} was introduced by \citet{AzLe19a}  and is known to satisfy \rpjrp \citep{BrPe23a}. Given a profile $R$ and a committee size $k$, it works as follows. First, we assume that each voter has a budget of $1$ and that each candidate has a cost of $\frac{n}{k+1}$. Further, EAR keeps track of a parameter $s$, which will measure the approval depth of the voters. Now, the rule works iteratively, starting with $s=1$. In each round, we do the following: for each non-selected candidate $x$, we let $b_s(x)$ denote the total budget of the voters who rank $x$ among their top-$s$ candidates. If there is a candidate $x$ such that $b_s(x)>\frac{n}{k+1}$, we add such a candidate to the committee and decrease the budget of the voters who put $x$ in their top-$s$ by $\frac{n}{k+1}$ in total. We repeat this addition step until no candidate satisfies the budget constraint anymore, in which case we increase the approval depth $s$ by $1$. This process is repeated until $k$ candidates are selected. Just as for STV, for fully defining EAR, it is necessary to say which candidate among the eligible ones should be selected and which voters pay for the candidate. However, this will not affect our result, so we do not specify these details. 
	
	\paragraph{Solid Coalition Refinement Rule.} The \emph{Solid Coalition Refinement Rule (SCR)} was recently suggested by \citet{ALP+25a} as a rule that satisfies \psc and committee monotonicity. To define this rule, we observe that a solid coalition can be seen as a pair of sets of voters and candidates $(S,X)$ such that $S\subseteq N$, $X\subseteq C$, and $X\succ_S C\setminus X$. Then, we define the underrepresentation value of a solid coalition $(S,X)$ by a committee $W$ by $\rho(W,S,X)=\frac{|S|}{1+|X\cap W|}$. Now, the solid coalition refinement rule builds the committee sequentially, starting with $W=\emptyset$. In each round, it first determines the solid coalition $(S^*,X^*)$ that maximizes $\rho(W,S,X)$ among all solid coalitions $(S,X)$ in the input profile with $X\not\subseteq W$ and $X \subsetneq C$. If $X^*\setminus W$ is a singleton, we add the corresponding candidate to the committee and proceed with the next round. Otherwise, for deciding which candidate from $X^*$ to select, a process of refinement is used: we repeatedly determine the solid coalition $(S',X')$ maximizing $\rho(W,S,X)$ with $X\not\subseteq W$ and $X\subsetneq X^*$ and update $(S^*,X^*)$ to $(S',X')$, until there is a single candidate in $X^*\setminus W$. We then add this candidate and proceed to the next round. This process is repeated $k$ times to select a committee of the given target size.\medskip
	
	We will next show that STV satisfies \spsc whereas EAR and SCR fail this condition. We believe this insight is interesting for at least two reasons. First, it means that the dual of STV always finds committees in the \pmvc, so our central veto proportionality notion is always satisfiable. Second, this result gives an argument for STV over EAR and SCR. Indeed, the latter two rules are typically motivated by the fact that they give stronger fairness guarantees or satisfy additional consistency notions. By contrast, the subsequent theorem shows that, in the sense of \spsc, STV gives a stronger fairness guarantee than the other two rules.  
	The proof of the following theorem can be found in Appendix~\ref{app:rules}.
	
	\begin{restatable}{theorem}{classicrules}\label{prop:classicrules}
		The following claims are true: 
		\begin{enumerate}[label=(\arabic*), leftmargin=*]
			\item STV satisfies \spsc. The dual of STV thus satisfies \pmvc.
			\item EAR and SCR fail \spsc. The duals of these rules thus fail \pmvc.
		\end{enumerate}
	\end{restatable}
	
	\subsection{Rules Inspired by \psvc}
	\label{subsec:newRules}
	
	In this section, we consider another natural  approach for designing voting rules that select committees in the \pmvc. Specifically, we will next show that, by repeatedly selecting the winners of singlewinner voting rules that find candidates in the proportional singlewinner veto core, we can construct a committee in the proportional multiwinner veto core. To make this formal, we will first introduce two (classes of) singlewinner voting rules, namely the Simultaneous Veto Rule by \citet{KiKe23a} and the Serial Veto Rule (aka Vote by Veto), which has been repeatedly rediscovered \citep{Moul82a,KiKe23a}.

	\paragraph{Simultaneous Veto.} In \emph{Simultaneous Veto}, the idea is that voters continuously veto their least-preferred candidate and the last fully vetoed candidate is the winner. To make this more precise, let $\mathtt{B}(\succ,X)$ denote the least preferred candidate in $X$ with respect to $\succ$. Moreover, we denote by $\Phi(x,t)$ the vetoes gathered for candidate $x$  by time $t$ and by $Y(t)$ the set of candidates that are not fully vetoed. At time $t=0$, it holds that $\Phi(x,t)=0$ for all $x\in C$ and $Y(t)=C$. Then, Simultaneous Veto works as follows: at every point in time $t$, $\Phi(x,t)$ increases at rate $|\{i\in N\colon \mathtt{B}(\succ_i, Y(t))=x\}|$. A candidate is fully vetoed once $\Phi(x,t)=1$ and we immediately remove such candidates from $Y(t)$. After removing a candidate $x$ from $Y(t)$, the voters who vetoed $x$ move on to their next least preferred candidate. This process is executed until, at time $m/n$, each candidate is vetoed, and we return the last fully vetoed candidate as the winner (with ties broken arbitrarily).
	
	\paragraph{Serial Veto.} The idea of \emph{Serial Veto} is similar to that of Simultaneous Veto, but voters now cast their vetoes sequentially. In more detail, for a given profile $R$, Serial Veto first picks a positive integer $t$ and a sequence $(v_1,\dots, v_{nmt})$ that contains each voter $mt$ times. We then iterate through the voters according to this sequence, and each voter vetoes her least preferred remaining candidate. To make this more formal, let $Y$ denote the currently available set of candidates. Then, the current voter $v_i$ increases the veto score of $\mathtt{B}(\succ_{v_i}, Y)$ by $1$ and, if this candidate has afterwards a veto score of $nt$, it is removed from $Y$. Just as for Simultaneous Veto, the winner is the last candidate whose veto score reaches $nt$. We emphasize that Serial Veto is a class of voting rules, since we can pick for each profile a different veto sequence, which may affect the outcome.\bigskip
	
	We note that both Simultaneous Veto and each Serial Veto rule always pick candidates in the proportional singlewinner veto core \citep{Moul82a,KiKe23a}. Moreover, while perhaps not clear from the definition, Simultaneous Veto is a Serial Veto rule. To see this, we can set $t=(n!)^m$ and consider any round-robin sequence of voters (i.e., we repeatedly cycle through the $n$ voters in a fixed order). This multiplier guarantees that a candidate can only ever be fully vetoed after the last voter that bottom-ranks it in the current cycle over the $n$ voters has cast its veto, so the order within a cycle does not matter (up to tie-breaking) and the serial vetoing becomes simultaneous.
	
	Given a singlewinner voting rule, perhaps the easiest way to extend it to a multiwinner voting rule is to repeatedly pick the winner according to the singlewinner rule, add it to the committee, and delete it from the profile. We call this the sequential extension of a singlewinner voting rule. To formally define this, we denote by $R|_X$ the restriction of a profile $R$ to the candidates $X$, i.e., $R|_X$ arises from $R$ by deleting all candidates outside of $X$ from the preference relations of the voters. Further, we view singlewinner voting rules as functions that map each profile to a singleton set of candidates. 
	Then, the \emph{sequential extension} of a singlewinner voting rule $f$ is defined recursively by $f^\mathit{seq}(R,1)=f(R)$ and $f^\mathit{seq}(R,k)=f^\mathit{seq}(R, k-1)\cup f(R|_{C\setminus f^\mathit{seq}(R, k-1)})$ for all $k\in \{2,\dots, m\}$.
	
	We next discuss an example to illustrate the Simultaneous Veto rule and its sequential extension.
	
	\begin{example}
		To illustrate how a singlewinner voting rule can be sequentially extended to the multiwinner setting, consider the following profile $R$ with $n=5$ voters and $m=4$ candidates, and let $k=2$. 
		\begin{center}
			\begin{tabular}{cc}
				3: &\, a, b, c, d \\
				1: &\, c, d, a, b \\
				1: &\, c, d, b, a 
			\end{tabular}
		\end{center}
		Consider the sequential extension of the Simultaneous Veto rule $f$, denoted by $f^{\mathit{seq}}$. In the first round, $f^{\mathit{seq}}(R,1)=f(R)$. Initially, three voters veto their bottom-ranked candidate $d$, while one voter vetoes $b$ and another vetoes $a$. Candidate $d$ is eliminated first at time $\frac{1}{3}$, after which the three voters turn to $c$, which is eliminated at time $\frac{2}{3}$. Candidate $b$ is then eliminated at time $\frac{3}{4}$, leaving $a$ as the last remaining candidate. Hence, $a$ is selected first. We then remove $a$ from $R$ and consider the restricted profile $R|_{C\setminus\{a\}}$. In the second round, $d$ is again eliminated first at time $\frac{1}{3}$, followed by candidate $b$ at time $\frac{1}{2}$. Thus, $f(R|_{C\setminus \{a\}})=\{c\}$, and consequently $f^{\mathit{seq}}(R,2)=\{a,c\}$. In contrast, one may directly apply Simultaneous Veto to $R$ and select the last $k$ candidates to be eliminated. This yields $\{a,b\}$, which differs from the outcome of the sequential extension. Moreover, one can verify that $\{a,c\}$ is in the \pmvc, whereas $\{a,b\}$ is not.
	\end{example}
	
	We will show next that, if the singlewinner voting rule $f$ always picks candidates from the proportional singlewinner veto core, its sequential extension always picks a committee from the proportional multiwinner veto core. Thus, the sequential extensions of Simultaneous Veto and all Serial Veto rules produce committees in the \pmvc. While this insight may not sound very impressive by itself, the duals of our sequential extensions satisfy \spsc and thus constitute new rules satisfying PSC. Given that only a few rules are known to satisfy this property, we believe that the subsequent result is thus rather useful. 
	
	\begin{theorem}\label{thm:general_seq_single_winner_veto_pmvc}
		A singlewinner voting rule is guaranteed to select a candidate in the proportional singlewinner veto core if and only if its sequential extension is guaranteed to select a committee in the proportional multiwinner veto core. 
	\end{theorem}
	
	\begin{proof}
		Assume $f$ is a singlewinner voting rule that always picks candidates from the proportional singlewinner veto core. We next show that its sequential extension $\Seqf$ satisfies \pmvc. We prove the statement by induction on the committee size. For the induction basis, let $k=1$. In this case, $\Seqf$ simply returns the candidate selected by $f$, which belongs to the proportional singlewinner veto core by assumption. Now, suppose that $f$ selects candidate $c$ when $k=1$ and it violates \pmvc. Then there exists a group of voters $S\subseteq N$, an integer $\ell$, and candidate sets $X,Y\subseteq C$ such that $|S| > \ell\frac{n}{m}$, $Y\succ_S X$, and $|\{c\}\cap X| > \max\{0, m-\ell - |Y|\}$. In particular, we have $c\in X$ and $1 > \max\{0, m-\ell-|Y|\}$. Since $m-\ell - |Y|$ is an integer, the inequality holds only if $m-\ell - |Y|\leq 0$, i.e., $\ell \geq m-|Y|$. We then rewrite the size of the group $|S| > \ell \frac{n}{m}\geq \frac{(m-|Y|)n}{m}$. Since each voter $i$ in $S$ prefers every candidate in $Y$ to $c$, the group $S$ blocks $c$ by the definition of proportional singlewinner veto core, a contradiction. Hence, $\Seqf$ satisfies \pmvc when $k=1$, thus proving our base case. 
		
		Now fix an integer $k > 1$ and assume inductively that the committee $W'$ selected by $\Seqf$ for target committee size $k-1$ satisfies \pmvc. Let $W$ denote the committee selected by $\Seqf$ for size $k$. By the definition of the sequential extension, there exists a candidate $x^*$ such that $W=W'\cup \{x^*\}$ and $\{x^*\}=f(R|_{C\setminus W'})$. Suppose, for contradiction, that $W$ fails \pmvc. Then there exists a group of voters $S\subseteq N$, an integer $\ell$, and candidate subsets $X,Y \subseteq C$ such that $|S| > \ell \frac{n}{m-k+1}$, $Y\succ_S X$, and $|W\cap X| > \max\{0, m-\ell-|Y|\}$. We first claim that $x^*\in X$. If this was not the case, it holds that $|W'\cap X|=|W\cap X| > \max\{0,m-\ell -|Y|\}$. Since $|S| > \ell \frac{n}{m-k+1} > \ell \frac{n}{m-k+2}$, this is a witness of a \pmvc violation by $W'$ for target committee size $k-1$, contradicting the induction hypothesis. Therefore, $x^*\in X$. 
		
		Next, let $r=|W\cap X|$. Since $x^*\in X$, there are exactly $r-1$ candidates in $X\cap W'$. Consider now the profile $R|_{C\setminus W'}$ for which $f$ selects $x^*$. For ease of exposition, we let $C'=C\setminus W'$ and note that $|C'|=m-k+1$. Furthermore, let $Y'=Y\cap C'$ be the candidates from $Y$ that are still available in this restricted election. We will next derive a lower bound on $|Y'|$, which, in particular, shows that this set is non-empty. To this end, we first observe that $W'$ contains $(k-1)-(r-1)=k-r$ candidates in $C\setminus X$ since $|X\cap W'|=r-1$ and $|W'|=k-1$. Moreover, we know $X\cap Y=\emptyset$. Hence, at most $k-r$ candidates from $Y$ belong to $W'$, which implies that $|Y'|\geq |Y|-(k-r)$. On the other hand, our contradiction assumption implies that $r > m-\ell-|Y|$, which further yields that $r \geq m-\ell-|Y|+1$ as all of the quantities are integers. By combining these insights, we obtain that
		\begin{align*}
			|Y'| \geq |Y|-(k-r) \geq |Y|-k + (m-\ell - |Y|+1)=m-k+1-\ell.
		\end{align*}
		
		Since, $|S|>\ell\frac{n}{m-k+1}$ by assumption, it follows that $\ell\leq (m-k)$. In combination with our last inequality, this shows that $|Y'|\geq 1$, i.e., $Y'$ is non-empty. Further, since $|C'|=m-k+1$, our inequality equivalently states that $|Y'|\geq |C'|-\ell$. This means that, in $R|_{C'}$, there are at most $\ell$ candidates outside of $Y' $. Furthermore, the size constraint on $S$ can be equivalently written as $\ell<|S|\cdot\frac{m-k+1}{n}=|S|\cdot\frac{|C'|}{n}$. Together, our two inequalities show that $|C'|-|Y'|\leq \ell < |C'|\frac{|S|}{n}$. Thus, group $S$ has sufficient veto power to veto all the candidates outside $Y'$ in $R|_{C'}$. Finally, since $x^*\in X$ and $Y\succ_S X$, it follows that $S$ and $Y'$ block $x^*$ by the definition of the proportional singlewinner veto core, which contradicts the assumption that $f$ always selects a candidate in the proportional singlewinner veto core. Therefore, we conclude that the committee $W$ returned by $\Seqf$ satisfies \pmvc.
		
		The converse direction is immediate as the \pmvc reduces to the \psvc and $\Seqf$ to $f$ when $k=1$.
	\end{proof}

\begin{remark}
	The dual rule of a sequential extension of a singlewinner voting rule works by using a singlewinner voting rule to repeatedly identify the worst candidate and deleting it from the profile. The winning committee of size $k$ then consists of the candidates that remain after $m-k$ such deletion steps. While giving a slightly different description, \citet{ALP+25a} have recently considered this class under the name ``reverse sequential rules" to construct voting rules that satisfy PSC and committee monotonicity. Notably, two of the rules explored by these authors, namely the reverse sequential Ordered Phragm\'en rule and the reverse sequential STV rule, are related to the rules discussed in this section. In particular, the reverse sequential Ordered Phragm\'en rule is the dual of the sequential extension of Simultaneous Veto, and reverse sequential STV is the dual of the sequential extension of a Serial Veto rule. 
\end{remark}

\begin{remark}\label{rem:PO}
	One may wonder whether the rules discussed in this section offer any advantage compared to STV and its dual. We believe that there are numerous reasons. First, it is easy to see that the sequential extension of each rule and its dual are committee monotone, which is not true for STV. Second, by construction, the dual of the sequential extension of every Serial Veto rule satisfies the following property: if $x$ is selected and $y$ is unselected then at least $\lceil\frac{n}{m}\rceil$ voters prefer $x$ to $y$. This can be seen as a stability notion, ensuring that each proposed swap between an unselected and selected candidate contradicts the preferences of a sizeable share of the population. On the other hand, for STV, it can happen that it selects a candidate that is almost Pareto-dominated: by the deletion dynamic, it can happen that STV selects a candidate even though all but $2$ voters prefer another unselected candidate to it. We refer to \Cref{app:Pareto} for details.
\end{remark}

\begin{remark}
	All sequential extensions and their duals fail \rpjrp. Indeed, by construction, these rules satisfy a property called committee monotonicity and \citet{ALP+25a} have shown that every rule that satisfies this condition fails \rpjrp.
\end{remark}

\subsection{Characterizing Sequential Rules satisfying \pmvc}
\label{subsec:characterization}

\Cref{thm:general_seq_single_winner_veto_pmvc} demonstrates that we can derive a committee in the proportional multiwinner veto core by repeatedly choosing from the singlewinner veto core. We will next turn to the converse question by asking which sequential committee voting rules satisfy \pmvc. As we will show, the only such rules will be the sequential extensions of Serial Veto rules. Moreover, we also present a characterization of the Simultaneous Veto rule to give a clear argument for which Serial Veto rule one might want to use for selecting from the proportional multiwinner veto core.

Now, to state our characterizations, we first need to formalize what we mean by a sequential rule. One previously discussed axiom that seems useful for this is committee monotonicity, which requires of a committee voting rule $f$ that $f(R,k)\subseteq f(R,k+1)$ for all profiles $R$ and committee sizes $k\in \{1,\dots, m-1\}$. Put differently, this axiom demands that all candidates that are chosen for a committee size $k$ must also be chosen for the committee size $k+1$. A consequence of this condition is that the committee can be built up by adding one candidate at a time. However, this axiom still leaves too much slack for a characterization, as it, e.g., allows that the choice of a candidate depends on which candidates have been chosen previously. We hence strengthen committee monotonicity by additionally requiring that the newly added candidate would also be selected by our rule if we ask it for a committee of size $1$ after deleting the previously selected candidates. We call the resulting notion committee expansion consistency.

\begin{definition}
	\label{def:committee_expansion_consistency}
	A committee voting rule $f$ satisfies committee expansion consistency if for any $0\leq k< m$, it holds that $f(R,k+1)=f(R, k) \cup f(R|_{C\setminus f(R,k)}, 1)$.
\end{definition}

To compare committee monotonicity to committee expansion consistency, it is illustrative to think about the sequence in which a rule picks the candidates. Then, committee monotonicity merely implies that such a sequence exists for every profile \citep{Elkind2017}. By contrast, committee expansion consistency additionally demands that this sequence is invariant under deleting an arbitrary prefix of it. This latter property is known as independence at the top in rank aggregation \citep{FBC14a,BBP26a}. 

We will next show that a committee voting rule satisfies \pmvc and committee expansion consistency if and only if it is a sequential extension of a Serial Veto rule. We note that the following result builds on a characterization of Serial Veto rules by \citet{KiKe23a} (see also \citet[Thm. 4-bis]{Moul83a}).

\begin{theorem}
	\label{thm:characterization_mpvc_consistency}
	A committee voting rule satisfies \pmvc and committee expansion consistency if and only if it is the sequential extension of a Serial Veto rule. 
\end{theorem}

\begin{proof}
	Suppose $f$ satisfies \pmvc and committee expansion consistency. We will first show that committee expansion consistency implies that there is a singlewinner voting rule $g$ such that $f=g^\mathit{seq}$. To this end, we set $g(R)=f(R,1)$ for every profile $R$. This immediately shows that $f(R,1)=g^\mathit{seq}(R,1)$ for every profile $R$. Now, inductively assume that $f(R,k)=g^{\mathit{seq}}(R,k)$ for every profile $R$ and some $k\in \{1,\dots, m-1\}$. By committee expansion consistency, we know that $f(R,k+1)=f(R,k)\cup f(R|_{C\setminus f(R,k)} ,1)$. By the induction hypothesis, we have that $f(R,k)=g^{\mathit{seq}}(R,k)$ and the definition of $g$ requires  that $f(R|_{C\setminus f(R,k)} ,1)=g(R|_{C\setminus f(R,k)})$. Thus, $f(R,k)\cup f(R|_{C\setminus f(R,k)} ,1) =g^{\mathit{seq}}(R,k)\cup g(R|_{C\setminus f(R,k)})$. Further, by the definition of sequential extensions, $g^{\mathit{seq}}(R,k)\cup g(R|_{C\setminus g^\mathit{seq}(R,k)})=g^{\mathit{seq}}(R,k+1)$. Chaining our equations then shows that $f(R,k+1)=g^{\mathit{seq}}(R,k+1)$, so $f$ is indeed the sequential extension of a singlewinner rule. 
	
	Next, since $f$ satisfies \pmvc, it follows that $g$ satisfies \psvc. Indeed, for $k=1$, the proportional multiwinner veto core reduces to the proportional singlewinner veto core and $g(R)=f(R,1)$ for every profile $R$. Next, a characterization by \citet{KiKe23a} shows that a candidate is in the singlewinner veto core if and only if there is a Serial Veto rule that selects this candidate. Hence, for every profile $R$, there exists a Serial Veto sequence whose selected candidate is the unique element of $g(R)$. Since the sequence may depend on the profile, $g$ is a Serial Veto rule, and therefore $f$ is its sequential extension.
	
	For the other direction, suppose that $f$ is the sequential extension of some Serial Veto rule $g$, i.e., $f=g^{\mathit{seq}}$. Since $g$ is a Serial Veto rule, it always selects a candidate in the \psvc. By \Cref{thm:general_seq_single_winner_veto_pmvc}, $g^{\mathit{seq}}$ satisfies \pmvc. Furthermore, by the definition of sequential extensions, we have $g^{\mathit{seq}}(R,k+1)=g^{\mathit{seq}}(R,k)\cup g(R|_{C\setminus g^{\mathit{seq}}(R,k)})$, which further implies $g^{\mathit{seq}}(R,k+1)=g^{\mathit{seq}}(R,k)\cup g^{\mathit{seq}}(R|_{C\setminus  g^{\mathit{seq}}(R,k)},1)$ as $g(R)=g^{\mathit{seq}}(R,1)$. This proves that $f=g^{\mathit{seq}}$ also satisfies committee expansion consistency.
\end{proof}

\Cref{thm:characterization_mpvc_consistency} gives a clear argument for when to use sequential extensions of Serial Veto rules. Moreover, we note that every multiwinner voting rule that always returns a committee in the proportional multiwinner veto core implies a singlewinner voting rule that always produces candidates in the proportional singlewinner veto core by restricting attention to committees of size $1$. This insight is interesting for two reasons: firstly, it means that, for $k=1$, the class of rules that always return candidates in the \pmvc is precisely the class of Serial Veto rules. Secondly, by taking the sequential extension of the induced singlewinner voting rules, we get another committee voting rule that satisfies \pmvc and committee expansion consistency. 

Motivated by \Cref{thm:characterization_mpvc_consistency}, a natural follow-up question is which Serial Veto rule one might want to use when constructing a multiwinner rule via the sequential extension. We believe that a natural requirement for this purpose is that no unanimous group of voters is advantaged by the chosen veto sequence. To formalize this, let $n(\succ)$ denote the number of voters who report a preference relation $\succ$ in the profile $R$, and recall that $n$ is the total number of voters. It then seems appealing to use veto sequences such that, for every preference relation $\succ$ and every position $\ell$ in the sequence, the number of vetoes cast by voters reporting $\succ$ is at most $\lceil \ell\frac{n(\succ)}{n}\rceil$. More formally, we say that a veto sequence $V=(v_1,\dots,v_{nmt})\in N^{nmt}$ is \emph{group-balanced} if, for every $\ell\in\{1,\dots,nmt\}$ and every preference relation $\succ$, it holds that $|\{\ell'\in\{1,\dots,\ell\}\colon {\succ_{v_{\ell'}}}={\succ}\}|\leq\lceil\ell\frac{n(\succ)}{n}\rceil$. Thus, no group of voters with identical preferences can exercise more than its proportional share of the veto power at any point in the veto process, ensuring that no such group is advantaged by being allowed to cast its vetoes at particularly convenient times. We further note that closely related balanced sequences of agents are commonly studied in allocation problems \citep[e.g.,][]{AWX15a,Aziz2020ExAnteExPost,AmanatidisEtAl2023FairDivision}, where they often guarantee fair outcomes. Furthermore, we define a Serial Veto rule to be group-balanced if it can always be implemented by a group-balanced veto sequence.

\begin{definition}[Group-Balancedness]
	A Serial Veto rule $f$ is \emph{group-balanced} if, for every profile $R$, there exists a group-balanced veto sequence $V$ such that $f(R)=g_V(R)$, where $g_V$ denotes the Serial Veto rule induced by $V$.
\end{definition}

As we will show next, this condition, together with the common side condition of homogeneity, forces us to use Simultaneous Veto, subject to some tie-breaking. To formalize this, let $W_{\mathrm{SV}}(R)$ denote the set of candidates that can be chosen by Simultaneous Veto under some tie-breaking at profile $R$. Moreover, a voting rule $f$ satisfies \emph{homogeneity} if it is invariant under replicating the full profile: $f(R)=f(\lambda R)$ for every profile $R$ and positive integer $\lambda\in\mathbb{N}$. Here, $\lambda R$ denotes the profile consisting of $\lambda$ copies of $R$. We show that a homogeneous Serial Veto rule is group-balanced if and only if it selects a candidate from $W_{\mathrm{SV}}(R)$ at every profile $R$. Together with \Cref{thm:characterization_mpvc_consistency}, this provides a strong argument for the sequential extension of Simultaneous Veto: \Cref{thm:characterization_mpvc_consistency} shows that we must use the sequential extension of a Serial Veto rule if we insist on \pmvc and committee expansion consistency, while the following result shows that, subject to homogeneity, only Simultaneous Veto can be implemented by an unbiased veto sequence. The proof of the following result is deferred to Appendix~\ref{app:characterizations}.

\begin{restatable}{theorem}{simultaneousVeto}\label{thm:simultaneousVeto}
	Let $f$ be a homogeneous Serial Veto rule. It holds that $f(R)\subseteq W_{\mathrm{SV}}(R)$ for every profile $R$ if and only if $f$ is group-balanced.
\end{restatable}

\begin{remark}
	\Cref{thm:characterization_mpvc_consistency} can be turned into a characterization of reverse sequential rules satisfying \spsc. To this end, we say a committee voting rule $f$ satisfies \emph{committee refinement consistency} if $f(R|_{f(R,k+1)},k)=f(R,k)$ for every profile $R$ and $k\in \{1,\dots, m-1\}$.  Now, it can be checked that, if $f$ satisfies \spsc and committee refinement consistency, its dual satisfies \pmvc and committee expansion consistency. By \Cref{thm:characterization_mpvc_consistency}, this means that $f$ is the dual rule of a sequential extension of a Serial Veto rule. These rules also have a more natural description: they sequentially delete candidates and, in each round, we consider a sequence of the voters $(v_1, \dots, v_{nmt})$ for some positive integer $t$. Then, we iterate through the voters, who vote for their top-ranked candidate that is still available. A candidate is secured if it gathered $nt$ votes and it is removed after this. Finally, the last candidate that is secured will be deleted from the profile. We believe this insight to be of independent interest as it is a constructive characterization of (reverse sequential) committee rules that satisfy \spsc. Furthermore, using this dual approach, the characterization of the sequential extension of Simultaneous Veto can be turned into a characterization of the reverse sequential extension of the Ordered Phragm\'en rule. 
\end{remark}

\subsection{Towards Improved Proportionality Guarantees}\label{subsec:SpscAndRpjrp}
As we have seen so far, committees satisfying \pmvc (and thus by duality also \spsc) are guaranteed to exist and allow for appealing structural insights. In this section, we are thus turning towards the question of whether one can give even stronger proportionality guarantees in committee voting with ranked preferences. Specifically, since it is known that the core may be empty in our setting, we ask whether there are committees that jointly satisfy \spsc and \rpjrp. By our duality framework, this is equivalent to the existence of a committee in the \pmvc that satisfies the dual of \rpjrp, but we find it more compelling to answer this question through the lens of inclusion-based proportionality. 

Now, we first recall that both of these axioms are logically incomparable (\Cref{prop:landscape}) and that EAR, the only rule known to satisfy \rpjrp, fails \spsc (\Cref{prop:classicrules}). Moreover, STV is known to fail \rpjrp and all rules in \Cref{subsec:newRules} fail this axiom as well. Despite all of this, we will show that committees satisfying \spsc and \rpjrp indeed always exist, thereby demonstrating that these two demanding fairness notions are jointly satisfiable. To prove this claim, we will use a ``repair" strategy: we first compute a committee satisfying \spsc and then repair possible violations of \rpjrp while maintaining \spsc. In more detail, to repair a \rpjrp violation, we compute an alternative winning committee by combining a set of candidates chosen for and funded by the underrepresented group with a smaller winning committee, which is inductively assumed to satisfy \spsc and \rpjrp. By formalizing this induction in terms of payment schemes and budgets, our theorem will follow.

\begin{theorem}
	\label{thm:spsc_rpjrp_compatibility}
	Given any profile $R$ and committee size $k$, there always exists a committee that satisfies both \spsc and \rpjrp.
\end{theorem}
\begin{proof}
	Fix an arbitrary profile $R$ with voter set $N$ and candidate set $C$. To show this theorem, we will first introduce some additional notation. First, for each voter $i\in N$ and threshold $r\in\{1,\ldots,m\}$, we denote by $\tp_i(r)=\{c\in C:\rank(c,\succ_i)\leq r\}$ the set of voter $i$'s top-$r$ candidates. Moreover, for every subset of voters $S\subseteq N$, we write $\tp_S(r)=\bigcup_{i\in S}\tp_i(r)$ and $\tp_{\emptyset}(r)=\emptyset$. Throughout this proof, we will think in terms of payments and budgets. Specifically, we suppose that each voter has an initial budget of $\frac{k+1}{n}$, and each selected candidate has a cost of $1$. This cost has to be paid by the voters when adding a candidate to the committee. We will frequently consider intermediate situations where some candidates have been added to the committee and the budgets of some voters have been reduced. To capture these situations, we define \emph{budget instances} as triples $(C', \kappa, b)$, where $C'\subseteq C$ is the set of available candidates, $\kappa\leq |C'|$ is the number of remaining seats in the committee, and $b=(b_i)_{i\in N}\in \mathbb{Q}^n_{\geq 0}$ is a rational vector specifying the remaining budget of every voter such that $\sum_{i\in N} b_i=\kappa+1$. Further, we let $b(S)=\sum_{i\in S} b_i$ for each budget vector $b$ and group of voters $S\subseteq N$.
	Given a budget instance $(C',\kappa,b)$, we say that a size-$\kappa$ committee $W\subseteq C'$ satisfies \spsc if, for every group of voters $S\subseteq N$, non-empty candidate set $X\subseteq C'$, and possibly empty set $Z\subseteq C'\setminus X$ with $C'\setminus (X\cup Z)\neq \emptyset$ and $X\succ_S C'\setminus (X\cup Z)$, it holds that $X\subseteq W$ or $b(S) \leq |Z|+|X\cap W|+1$.
	Similarly, a size-$\kappa$ committee $W\subseteq C'$ satisfies \rpjrp for a budget instance $(C', \kappa, b)$ if, for every non-empty set $S\subseteq N$, threshold $r\in \{1,\dots, m\}$, and candidate $c\in (C'\setminus W)\cap \bigcap_{i\in S}\tp_i(r)$, we have $b(S) \leq 1 + |W\cap \tp_S(r)|$. We emphasize that, for the definition of \rpjrp, $\tp_S(r)$ still refers to the union of the top-$r$ candidates of the voters $i\in S$ when all candidates are available. By contrast, for a \rpjrp violation, the candidate witness $c$ has to come from $C'\setminus W$. Furthermore, \spsc is only concerned with candidates in $C'$. Finally, for $r\in\{1,\dots,m+1\}$, we say that a size-$\kappa$ committee $W\subseteq C'$ satisfies \rpjrp \emph{for all thresholds $t\geq r$} in a budget instance $(C', \kappa, b)$ if the \rpjrp condition above holds for all thresholds $t\in\{r,\dots,m\}$. For $r=1$, this is \rpjrp, and for $r=m+1$, the condition is vacuous.
	
	Now, our goal is to show that there is a committee $W$ that satisfies both \spsc and \rpjrp for the budget instance $(C, k, (\frac{k+1}{n})_{i\in N})$, which is equivalent to satisfying these axioms in the regular sense. To this end, we will prove that, for every $r\in\{1,\dots,m+1\}$ and every budget instance $(C',\kappa,b)$ with $0\leq \kappa \leq \min\{k,|C'|\}$, there is a size-$\kappa$ committee $W$ that satisfies \spsc and \rpjrp for all thresholds $t\geq r$. Applying the claim with $r=1$ to the instance $(C,k,(\frac{k+1}{n})_{i\in N})$ proves the theorem. To prove our claim, we use an induction on $\kappa$ and, for each fixed $\kappa$, a backward induction on $r$. Specifically, we fix two values $\kappa\in \{0,\dots, k\}$ and $r\in \{1,\dots, m+1\}$ and assume that \emph{(i)} for every budget instance $(C',\kappa',b)$ with $\kappa'<\kappa$, there exists a size-$\kappa'$ committee satisfying \spsc and \rpjrp, and \emph{(ii)} if $r\leq m$, then, for every budget instance $(C',\kappa, b)$, there exists a size-$\kappa$ committee satisfying \spsc and \rpjrp for all thresholds $t\geq r+1$.
	
	For the base case of the outer induction, $\kappa=0$, the claim is trivial since the empty set satisfies both \spsc and \rpjrp. Hence, we assume that $\kappa\geq 1$. For the inner induction, the base case is $r=m+1$. In this case, \rpjrp is vacuous, so we only need to find a committee satisfying \spsc for the considered budget instance $(C',\kappa,b)$. When $\kappa=|C'|$, $W=C'$ satisfies \spsc since $X\subseteq W$ for all $X\subseteq C'$. On the other hand, when $0<\kappa<|C'|$, consider a positive integer $\lambda$ such that $\lambda b_i\in\mathbb{N}$ for all $i\in N$. Such a scalar exists as $b$ is rational. Let $R'$ be the profile on $C'$ with voter set $N'$ in which, for every $i\in N$, there are $\lambda b_i$ voters reporting the preference relation of $i$ restricted to $C'$. In particular, this means that $|N'|=\lambda(\kappa+1)$. Let $W$ be the committee returned by STV for $R'$ and committee size $\kappa$. By Claim (1) in \Cref{prop:classicrules}, $W$ satisfies the non-budgeted definition of \spsc. Thus, for every set of voters $S'\subseteq N'$, disjoint candidate sets $X,Y\subseteq C'$ with $X\succ_{S'}Y$, and integer $\ell\in\mathbb{N}$ with $|S'|>(\ell+|C'\setminus(X\cup Y)|)\frac{|N'|}{\kappa+1}$, we have $|W\cap X|\geq\min(|X|,\ell)$. Now suppose $W$ violates \spsc for $(C',\kappa,b)$. Then there is a set of voters $S\subseteq N$, a non-empty set $X\subseteq C'$, and a possibly empty set $Z\subseteq C'\setminus X$ with $X\succ_S C'\setminus(X\cup Z)$, $X\not\subseteq W$, and $b(S)>|Z|+|X\cap W|+1$. Let $Y=C'\setminus(X\cup Z)$, $\ell=|X\cap W|+1\leq |X|$, and let $S'$ consist of all copies of the voters $i\in S$ in $N'$. Then $|S'|=\lambda b(S)>\lambda(\ell+|Z|)=(\ell+|C'\setminus(X\cup Y)|)\frac{|N'|}{\kappa+1}$, where the final equality uses $|N'|=\lambda(\kappa+1)$. This proves that $W$ fails \spsc for $R'$ since $\min(|X|,\ell)=\ell>|W\cap X|$, a contradiction.
	
	Finally, assume that $\kappa\geq 1$ and $r\leq m$, and that our induction hypothesis holds for all $\kappa'<\kappa$ and all $r'>r$ when $\kappa'=\kappa$. Fix a budget instance $(C',\kappa,b)$. By the induction hypothesis for $r+1$ and $\kappa$, there is a size-$\kappa$ committee $W$ that satisfies \spsc and \rpjrp for all thresholds $t\geq r+1$ in $(C',\kappa,b)$. If $W$ also satisfies \rpjrp for threshold $r$, then $W$ proves the claim for $r$ and $\kappa$ as well. Hence, assume that this is not the case. Then, there is a group of voters $S\subseteq N$ and a candidate $c\in (C'\setminus W)\cap\bigcap_{i\in S}\tp_i(r)$ with $b(S)>1+|W\cap\tp_S(r)|$. For fixed $c$ and $r$, we choose $S$ inclusion-minimal with this property, i.e., $b(S')\leq 1+|W\cap\tp_{S'}(r)|$ for every non-empty $S'\subsetneq S$. Furthermore, let $T=\{c\}\cup(W\cap\tp_S(r))\subseteq C'$, and define the payments $p_i$ by $p_i=|T|\frac{b_i}{b(S)}$ for all $i\in S$ and $p_i=0$ for all $i\notin S$. As usual, $p(U)=\sum_{i\in U}p_i$ for all $U\subseteq N$.
	
	We will next analyze the payments $p_i$ in more detail. First, since $c\in\tp_S(r)\setminus W$, we have $T\subseteq\tp_S(r)$ and $|T|=1+|W\cap\tp_S(r)|<b(S)\leq b(N)=\kappa+1$. Hence, $1\leq|T|\leq\kappa$ and $p_i\leq b_i$ for all $i\in N$. Moreover, $p(N)=|T|$ by definition. We next show that
	\begin{equation*}		
		p(U)\leq|T\cap\tp_{U\cap S}(r)|\quad\text{for all }U\subseteq N.\tag{$\ast$}
	\end{equation*}
	For this, let $S'=U\cap S$ and note that $p(U)=p(S')$. If $S'=\emptyset$, then $p(U)=0$. If $S'=S$, then $p(U)=|T|=|T\cap\tp_S(r)|$ as $T\subseteq\tp_S(r)$. Finally, if $\emptyset\neq S'\subsetneq S$, then $p(S')\leq b(S')\leq 1+|W\cap\tp_{S'}(r)|=|T\cap\tp_{S'}(r)|$, where the second inequality uses the minimality of $S$ and the last equality follows from $c\in\tp_{S'}(r)\setminus W$ and $\tp_{S'}(r)\subseteq\tp_S(r)$.
	
	Now, let $C''=C'\setminus T$, $\kappa''=\kappa-|T|$, and $b''_i=b_i-p_i$ for all $i\in N$. By the preceding observations, $b''\in\mathbb{Q}^n_{\geq0}$, $b''(N)=\kappa+1-|T|=\kappa''+1$, and $0\leq\kappa''\leq|C''|$. Thus, $(C'',\kappa'',b'')$ is a budget instance with $\kappa''<\kappa$. By the induction hypothesis for $\kappa''$, there is a size-$\kappa''$ committee $W^c\subseteq C''$ that satisfies \spsc and \rpjrp for $(C'',\kappa'',b'')$. Let $W'=T\cup W^c$. Since $T\cap W^c=\emptyset$, $W'$ has size $\kappa$. It remains to show that $W'$ satisfies \rpjrp for all thresholds $t\geq r$ and \spsc for $(C',\kappa,b)$.
	
	For the claim on \rpjrp, consider a non-empty set of voters $U\subseteq N$, a threshold $t\in\{r,\dots,m\}$, and a candidate $c'\in(C'\setminus W')\cap\bigcap_{i\in U}\tp_i(t)$. Since $c'\in C'\setminus W'=C''\setminus W^c$, the \rpjrp guarantee of $W^c$ in $(C'', \kappa'', b'')$ gives $b''(U)\leq1+|W^c\cap\tp_U(t)|$. Since $r\leq t$ and $U\cap S\subseteq U$, we have $\tp_{U\cap S}(r)\subseteq\tp_{U}(r)\subseteq \tp_U(t)$. This implies that $|T\cap\tp_{U\cap S}(r)|\leq |T\cap \tp_U(t)|$. Together with $(\ast)$, it follows that
	\begin{align*}
		b(U)&=b''(U)+p(U)\\
		&\leq 1+|W^c\cap\tp_U(t)|+|T\cap\tp_{U\cap S}(r)|\\
		&\leq 1+|W^c\cap\tp_U(t)|+|T\cap\tp_{U}(t)|\\
		&= 1+|W'\cap \tp_U(t)|.
	\end{align*}
	This proves that $W'$ satisfies \rpjrp for all thresholds $t\geq r$ in $(C',\kappa,b)$.
	
	Finally, we turn to \spsc. For this, we fix a non-empty set of voters $U\subseteq N$, a non-empty candidate set $X\subseteq C'$, and another possibly empty set $Z\subseteq C'\setminus X$ such that $C'\setminus (X\cup Z)\neq\emptyset$ and $X\succ_U C'\setminus(X\cup Z)$. If $X\subseteq W'$, \spsc is trivially satisfied for $U$ with respect to $X$ and $Z$, so we suppose that $X\not\subseteq W'$. We need to show that $b(U)\leq|Z|+|X\cap W'|+1$. First, if $X\subseteq\tp_S(r)$, then $X\cap W\subseteq W'$ since $X\cap W\subseteq \tp_S(r)\cap W\subseteq T$. In particular, since $X\not\subseteq W'$, this means that $X\not\subseteq W$. Since $W$ satisfies \spsc for $(C',\kappa, b)$, we derive that $b(U)\leq |Z|+|X\cap W|+1\leq |Z|+|X\cap W'|+1$, so $W'$ satisfies \spsc in this case. Next, suppose that $X\not\subseteq\tp_S(r)$. In this case, we claim that $T\cap\tp_{U\cap S}(r)\subseteq X\cup Z$. If $U\cap S=\emptyset$, this follows immediately as $\tp_{U\cap S}(r)=\emptyset$. Otherwise, choose a candidate $y\in X\setminus\tp_S(r)$. Let $i\in U\cap S$ and consider a candidate $a\in T\cap\tp_i(r)$. Suppose that $a\notin X\cup Z$. Since $a\in C'\setminus(X\cup Z)$ and $y\in X$, it holds that $y\succ_i a$. However, because $a\in\tp_i(r)$, this implies that $y\in\tp_i(r)\subseteq\tp_S(r)$, a contradiction to the choice of $y$. Thus, $T\cap\tp_i(r)\subseteq X\cup Z$ for every $i\in U\cap S$, proving the claim. By $(\ast)$, this means that $p(U)\leq|T\cap \tp_{U\cap S}(r)|\leq |T\cap(X\cup Z)|=|T\cap X|+|T\cap Z|$. Now, let $X''=X\setminus T$ and $Z''=Z\setminus T$. These sets satisfy that $X''\neq\emptyset$ (as $X\not\subseteq W'$ and $T\subseteq W'$) and $Z''\subseteq C''\setminus X''$. Further, if $C''\setminus (X''\cup Z'')\neq\emptyset$, then $X''\succ_U C''\setminus(X''\cup Z'')$ since $C''\setminus(X''\cup Z'')\subseteq C'\setminus(X\cup Z)$. Moreover, $X''\not\subseteq W^c$ because otherwise $X\subseteq T\cup W^c=W'$. Thus, in this case, \spsc of $W^c$ yields $b''(U)\leq|Z''|+|X''\cap W^c|+1$. On the other hand, if  $C''\setminus (X''\cup Z'')=\emptyset$, then $C''=X''\cup Z''$. In this case, we again derive our inequality since $b''(U)\leq b''(N)=\kappa''+1=|W^c|+1\leq |Z''|+|X''\cap W^c|+1$. By combining these insights and using that $W^c$ and $T$ are disjoint, we now derive that
	\begin{align*}
		b(U)&=b''(U)+p(U)\\
		&\leq |Z\setminus T|+|X''\cap W^c|+1+|T\cap X|+|T\cap Z|\\
		&=|Z|+|X\cap W'|+1.
	\end{align*}
	Hence, we conclude that $W'$ satisfies \spsc in $(C',\kappa,b)$, which completes the induction step.
\end{proof}

\begin{remark}
	The proof of \Cref{thm:spsc_rpjrp_compatibility} yields a constructive way to compute a committee that satisfies both \spsc and \rpjrp: we can first compute a committee returned by STV and then recursively repair this committee so that it satisfies \rpjrp while maintaining \spsc. However, in its direct form, this repair approach does not yield a polynomial-time algorithm. Indeed, for each remaining committee size $\kappa\in\{1,\dots,k\}$, the approach may make up to $m$ recursive calls, each of which strictly decreases the remaining committee size. A straightforward analysis therefore gives a bound of $m^{O(k)}$ on the number of recursive calls. While a more careful analysis shows that fewer calls suffice, it remains open whether the repair scheme can be implemented with a polynomial number of recursive calls. Moreover, although our existence proof is constructive, it does not yield a particularly compelling rule. It therefore seems interesting to find a natural polynomial-time rule that always returns a committee satisfying both \spsc and \rpjrp.
\end{remark}

\begin{remark}
	The proof of \Cref{thm:spsc_rpjrp_compatibility} yields new insights into \rpjrp, as our repair scheme itself works also without \spsc. In this sense, this gives new approaches towards computing committees satisfying this property. Moreover, we note that the repair approach in the proof of \Cref{thm:spsc_rpjrp_compatibility} is conceptually reminiscent of works on committee reconfiguration \citep{chenEtAl2024,dongEtAl2026}. This literature studies whether one committee can be transformed into another through a sequence of local changes, typically single-candidate replacements, while maintaining a proportionality guarantee at every intermediate committee. Our construction can be viewed as a coarser form of such a reconfiguration process, since we make repeated local modifications to arrive at a committee satisfying both \spsc and \rpjrp. Unlike the standard reconfiguration model, however, an individual repair in our process may replace several candidates simultaneously. It would therefore be interesting to determine whether these repairs can always be decomposed into local changes that preserve \spsc throughout. More broadly, one may ask whether the committees satisfying \spsc, or those satisfying both \spsc and \rpjrp, form a connected reconfiguration space under single-candidate replacements.
\end{remark}

\section{Verifying the Proportional Multiwinner Veto Core}\label{sec:verification}

As our final contribution, we consider the computational complexity of deciding whether a given committee satisfies \pmvc or \spsc. This verification problem is important, for example, when running simulations that measure how frequently a voting rule fails these axioms or when auditing whether a chosen committee satisfies one of these fairness notions in practice. Unfortunately, we show next that the verification problems for \pmvc and \spsc are coNP-complete and thus computationally intractable unless $\text{P}=\text{coNP}$. This contrasts with existing fairness notions such as \psc and \rpjrp, both of which can be verified in polynomial time~\citep{BrPe23a}. Before establishing our hardness result, we present a useful characterization of violations of \pmvc and \spsc. We defer the proof of the following proposition and all other results in this section to Appendix~\ref{app:verification}.

\begin{restatable}{proposition}{violation}\label{lem:violation_witness_strong_psc}
	The following claims are true.
	\begin{enumerate}[label=(\arabic*), leftmargin=*]
		\item A size-$k$ committee $W$ violates \spsc for a profile $R$ if and only if there exist two non-empty sets $X,Y\subseteq C$ such that $|\{i\in N: X\succ_i Y\}|>(|C\setminus(X\cup Y)|+1)\frac{n}{k+1}$ and $X\cap W=\emptyset$.
		\item A size-$k$ committee $W$ is not in the proportional multiwinner veto core for a profile $R$ if and only if there exist two non-empty sets $X,Y\subseteq C$ such that $|\{i\in N: Y\succ_i X\}|>(|C\setminus(X\cup Y)|+1)\frac{n}{m-k+1}$ and $X\subseteq W$.
	\end{enumerate}
\end{restatable}

More intuitively, \Cref{lem:violation_witness_strong_psc} shows that if a committee fails \spsc, then there exist two witness sets $X$ and $Y$ such that $X$ is disjoint from the committee and enough voters prefer all candidates in $X$ to those in $Y$, so that they can demand that one candidate from $X$ be chosen. Analogously, if a committee fails \pmvc, there is a subset $X$ of the committee that witnesses this violation. While these insights provide additional structure for our fairness axioms, they do not imply that the axioms can be verified in polynomial time. Indeed, we show next that deciding whether a committee satisfies \spsc or \pmvc is coNP-complete.

\begin{restatable}{theorem}{verification}\label{thm:verification}
	Deciding whether a committee belongs to the \pmvc for a profile is coNP-complete. Analogously, deciding whether a committee satisfies \spsc for a profile is coNP-complete.
\end{restatable}

While \Cref{thm:verification} is a negative result for our new proportionality notions, we next show that, at least for small parameter values, they can still be efficiently verified, as they admit FPT algorithms in $m$ and $n$. More specifically, we establish the following two claims. First, we prove that, for a given set of candidates $X$, it can be checked in polynomial time whether \pmvc is violated for $X$, in the sense that we choose too many candidates from this set. In combination with Claim (2) of \Cref{lem:violation_witness_strong_psc}, this insight even yields an FPT algorithm parameterized by $k$, as we can check every subset of the committee to verify that it satisfies \pmvc. Second, we prove that, for a given set of voters $S$, it can be checked whether too many candidates from a set vetoed by $S$ are chosen. This yields an FPT algorithm parameterized by $n$ by iterating over all subsets of voters. For auditing purposes, these insights may be sufficient in practice, as we are usually interested in whether particular groups of voters or candidates are fairly represented. Moreover, while we phrase the subsequent proposition for \pmvc, the arguments can easily be adapted to \spsc.

\begin{restatable}{proposition}{fpt}\label{prop:FPT}
	The following claims are true.
	\begin{enumerate}[label=(\arabic*), leftmargin=*]
		\item Given a size-$k$ committee $W$, a profile $R$, and a non-empty set of candidates $X$, it can be checked in polynomial time whether $W$ fails \pmvc for $X$, i.e., whether there exist a non-empty set of candidates $Y$, a set of voters $S$, and an integer $\ell$ such that $Y\succ_S X$, $|S|>(\ell+|C\setminus(X\cup Y)|)\frac{n}{m-k+1}$, and $|X\setminus W|<\min(|X|,\ell)$.
		\item Given a size-$k$ committee $W$, a profile $R$, and a set of voters $S$, it can be checked in polynomial time whether $W$ fails \pmvc for $S$, i.e., whether there exist non-empty sets of candidates $X,Y$ and an integer $\ell$ such that $Y\succ_S X$, $|S|>(\ell+|C\setminus(X\cup Y)|)\frac{n}{m-k+1}$, and $|X\setminus W|<\min(|X|,\ell)$.
	\end{enumerate}
\end{restatable}

\section{Conclusion}

In this paper we study the concept of veto proportionality in multiwinner voting, which intuitively demands that, for each sufficiently cohesive group of voters, not too many strongly disliked candidates are included in the selected committee. Specifically, our central fairness notion is the proportional multiwinner veto core (\pmvc), which extends the proportional singlewinner veto core by \citet{Mou81a} to committee elections. We first show that the proportional multiwinner veto core naturally relates to existing proportionality notions in committee voting, such as proportionality for solid coalitions (\psc) and core-stability, through a duality framework. In particular, this duality framework maps classical inclusion-based proportionality notions to veto proportionality notions and vice versa, and we show that the dual axiom of the proportional multiwinner veto core, which we call \spsc, implies \psc and is implied by a variant of core-stability. Second, we turn to the question of how to find committees in the proportional multiwinner veto core. To this end, we show that STV, a well-known committee voting rule, satisfies \spsc, whereas other established rules, such as EAR \citep{AzLe19a} and SCR \citep{ALP+25a}, fail this condition. This means that the dual rule of STV returns committees in the \pmvc. We also show that by repeatedly applying singlewinner voting rules that select from the proportional singlewinner veto core, we can also compute committees in the \pmvc. Even more, under a strong form of committee monotonicity, we characterize the resulting sequential rules as the only ones that satisfy \pmvc. By the duality framework, this also yields a natural and novel class of committee rules that satisfy \spsc, which we believe to be of independent interest. Third, in a pursuit of stronger fairness guarantees, we show that \rpjrp and \spsc, which are logically incomparable but two of the most demanding fairness notions in committee voting that are always satisfiable, can be jointly satisfied by a single committee in every profile. This strengthens the fairness guarantees that are known to be always satisfiable. Finally, we also consider the computational problem of deciding whether a committee satisfies \pmvc or \spsc and show that these problems are coNP-complete. 

More broadly, our paper offers a new dual perspective on what proportionality means in committee elections, and we believe that there are numerous natural follow-up questions. For example, in many settings, it seems natural to balance inclusion and exclusion guarantees: if many voters like a candidate, we should include it, and if many voters dislike a candidate, we should exclude it. Unfortunately, naive adaptations of inclusion-based proportionality axioms and our veto-based proportionality axioms are easily seen to be jointly incompatible, since one group may insist on choosing a candidate that another group insists on excluding. The question is therefore whether approximate variants of such axioms can be jointly satisfied and to analyze this tradeoff in more depth. Moreover, our work also leaves some immediate questions open. For instance, one may try to construct a compelling polynomial-time rule that always returns committees satisfying \rpjrp and \spsc, unify these axioms into a natural, always satisfiable fairness notion, or investigate in more depth what can be achieved by sequential extensions of Serial Veto rules.  

\section*{Funding and Personal Acknowledgements}
Haris Aziz is supported by the NSF-CSIRO grant on ``Fair Sequential Collective Decision Making''. Patrick Lederer is supported by the European Research Council (ERC) via the ERC Synergy Grant on ``Advancing Digital Democratic Innovation'' (\url{doi.org/10.3030/101166894}). Funded by the European Union. Views and opinions expressed are, however, those of the authors only and do not necessarily reflect those of the European Union or the European Research Council Executive Agency. Neither the European Union nor the granting authority can be held responsible for them.

\section*{AI Acknowledgements}

We used LLMs, in particular ChatGPT-5.6 Sol and Claude Opus-5.5, to improve the exposition of this manuscript and to assist in checking proofs. Moreover, ChatGPT-6 Astra generated an initial proof of \Cref{thm:spsc_rpjrp_compatibility}. The proof presented in the manuscript was subsequently substantially simplified, reorganized, and independently verified by the authors. The authors take full responsibility for all results and arguments.

\bibliographystyle{abbrvnat} 
\bibliography{veto-MWV}

\appendix
\crefalias{section}{appendix}

\section{Omitted Proofs from Section 3}
\label{app:DualImplication}

\DualImplication*
\begin{proof}
	We will show here only the first claim. The second claim follows by observing that, in our framework, a committee voting rule can be seen as a fairness notion that always returns a singleton set. Now, for Claim (1), fix two fairness notions $G$ and $F$. First, assume that $G$ implies $F$, which means that $G(R,k)\subseteq F(R,k)$ for all profiles $R$ and committee sizes $k$. Furthermore, let $R$ be a profile, $k$ a committee size, and $W\in G^d(R,k)$ a committee satisfying the dual fairness notion of $G$. By the definition of the dual, we know that $C\setminus W\in G(\rev(R), m-k)$. By our set-inclusion, this means that $C\setminus W\in F(\rev(R), m-k)$. Finally, using the definition of the dual again, $C\setminus W\in F(\rev(R), m-k)$ implies that $W\in F^d(R,k)$. Since $W\in G^d(R,k)$ was chosen arbitrarily, this shows that $G^d(R,k)\subseteq F^d(R,k)$. The same argument essentially also applies for the converse direction of the equivalence, so we leave it to the reader. 
\end{proof}

\landscape*
\begin{proof}
	We split our proposition into four unrelated propositions: first, we show that core implies \spsc and \rpjrp, and that \spsc implies \psc. Then, we prove that \spsc and \rpjrp are logically unrelated. Finally, we recall that \citet{BrPe23a} have already shown that \rpjrp implies \psc (and that their proof also works under the Droop-quota).\medskip 
	
	\textbf{Core-stability implies \spsc.} Fix a profile $R$ and a committee $W$ of size $k$. Assume that $W$ fails \spsc. We next prove that it also fails core-stability. Since $W$ fails \spsc, there is a set of voters $S$, an integer $\ell$ and two candidate sets $X$, $Y$ such that $|S|>(\ell+|C\setminus (X\cup Y)|)\frac{n}{k+1}$, $X\succ_S Y$, and $|W\cap X|<\min(|X|, \ell)$. Let $Z=C\setminus(X\cup Y)$ and $\ell'=\ell+|Z|$. Using these definitions, we have that $|W\cap X|\leq \ell - 1=\ell'-|Z|-1$ and thus $\ell'\geq |Z|+ |W\cap X|+1$. Next, we observe that $X\setminus W \neq \emptyset$ because otherwise $X\subseteq W$ and thus $|W\cap X|=|X|\geq \min(|X|, \ell)$. We next construct a candidate subset $X'$ where $X'$ contains all candidates in $W\cap Z$, all candidates in $W\cap X$, and one candidate $c^*$ in $X\setminus W$, i.e., $X'=(W\cap X) \cup (W\cap Z) \cup \{c^*\}$. 
	We claim that $X'\setminus W\succ_S W\setminus X'$. To see this, we first note that $X'\setminus W=\{c^*\}$. Furthermore, since $X\cup Y\cup Z=C$, we know that $W\setminus X'\subseteq Y$. Also, we note that $|X'|\leq \ell'$ and that $|S|>\ell'\cdot \frac{n}{k+1}$. This means that $\ell'\leq k$, so we conclude that $W\setminus X'\neq\emptyset$. Finally, because $c^*\in X$, every voter in $S$ prefers $c^*$ to every candidate in $Y$ and thus $X'\setminus W\succ_S W\setminus X'$. This shows that $W$ also fails core-stability.\medskip
	
	\textbf{Core-stability implies \rpjrp.} Fix a profile $R$ and a committee $W$ of size $k$. We assume that $W$ fails \rpjrp and prove that it also fails core-stability. If $W$ fails \rpjrp, there is a group of voters $S$, a rank $r$, and an integer $\ell$ such that $|S|>\ell \frac{n}{k+1}$, $|W\cap \{x\in C\colon \exists i\in S\colon \text{rank}(x,\succ_i)\leq r\}|<\ell$, and there is a candidate $x^*\in C\setminus W$ such that $\rank(x^*,\succ_i)\leq r$ for all $i\in S$. Let $X'=(W\cap \{x\in C\colon \exists i\in S\colon \rank(x,\succ_i)\leq r\})\cup \{x^*\}$. By assumption $|X'|\leq \ell$ and the size constraint on $S$ implies that $\ell\leq k$. In combination with $x^*\in X'\setminus W$, this suffices to conclude that $W\setminus X'\neq\emptyset$. Further, it holds for all $y\in W\setminus X'$, $z\in X'\setminus W$ that $z\succ_i y$ for all $i\in S$, because $X'\setminus W=\{x^*\}$ and $\rank(y,\succ_i)>r$ for all $y\in W\setminus X'$. This proves that $W$ fails core-stability.\medskip
	
	\textbf{\spsc implies PSC.} Consider a profile $R$ and a feasible committee $W$ of size $k$. We assume that $W$ fails PSC and show that it also fails \spsc. Since $W$ fails PSC, we know that there exists an integer $\ell$, a subset of voters $S \subseteq N$ with $|S|>\ell \frac{n}{k+1}$, and a candidate set $X\subseteq C$ such that $X\succ_S C\setminus X$ and $|W\cap X| < \min\{|X|, \ell\}$. This implies that $W$ also violates \spsc. To see this, we can simply choose $S$ and $X$ as before and set $Y=C\setminus X$. In this case, $C\setminus (X\cup Y)=\emptyset$, so the size constraint on $S$ exactly reduces to the one of PSC. We further note that it is easy to see that PSC does not imply \spsc. \medskip
	
	\textbf{\rpjrp and \spsc are logically incomparable.} This result will be a direct consequence from some of our later work. In particular, in  \Cref{sec:rules}, we show that, e.g., STV and several committee monotone committee voting rules satisfy \spsc, but these rules are known to fail \rpjrp \citep{BrPe23a,ALP+25a}. This suffices to show that \spsc does not imply \rpjrp. Conversely, we show in Section 4 also that a rule called Expanding Approvals Rule fails \spsc, but this rule is known to satisfy \rpjrp~\citep{BrPe23a}. Hence, \rpjrp also does not imply \spsc. 
\end{proof}

\section{Omitted Proofs from Section 4}\label{app:satisfaction}

\subsection{Omitted Proofs from Section 4.1}\label{app:rules}
\classicrules*
\begin{proof}
	We prove the claims of this theorem separately and start with the counterexamples for EAR and SCR.\medskip
	
	\textbf{EAR fails \spsc.} Consider the following profile and let $k=6$. The $\dots$ symbol at the end of the voters' preference relations indicates that the remaining candidates can be ordered arbitrarily. 
	\begin{align*}
		&5: z_1,z_2,x_1,x_2,x_3, y_2,\dots\\
		&5: x_1,x_2,x_3,y_1,z_1, z_2, \dots\\
		&1: y_1, w_{1}^1, w_{2}^1, w_{3}^1, w_{4}^1, w_{5}^1,\dots\\
		&1: y_2, w_{1}^2, w_{2}^2, w_{3}^2, w_{4}^2, w_{5}^2,\dots \\
		&1: y_3, w_{1}^3, w_{2}^3, w_{3}^3, w_{4}^3, w_{5}^3,\dots. 
	\end{align*}
	
	We start by computing EAR for our profile. To this end, we observe that the cost of each candidate is $\frac{n}{k+1}=\frac{13}{7}$. Hence, for $r=1$ and $r=2$, we buy $x_1$, $x_2$, $z_1$ and $z_2$. After this, the budgets of our first two groups of voters are reduced to $5-\frac{26}{7}=\frac{9}{7}$. Hence, for $r=3$, no candidate is affordable. For $r=4$, $y_1$ is bought, as it is supported by a group with a budget of $\frac{9}{7}$ and by a single voter with budget $1$. For $r=5$, again no candidate can be bought. In particular, we note here that the budget for $x_3$ is at most $\frac{9}{7}$ from the first five voters and $\frac{3}{7}$ from the second group of five voters. Finally, for $r=6$, $y_2$ is bought by the first $5$ voters and the voter in the fourth line. Hence, EAR chooses $\{x_1,x_2,y_1,y_2,z_1,z_2\}$.
	
	We claim that this committee fails \spsc. To this end, let $S$ denote the group consisting of the first $10$ voters, let $X=\{x_1,x_2,x_3\}$, and $Y=C\setminus (X\cup \{z_1,z_2\})$. It holds that $|S|=10>5\cdot \frac{n}{k+1}$, so our group is large enough to demand $5$ candidates. Hence, \spsc requires that $|W\cap X|\geq \min(|X|, \ell-(m-|X|-|Y|))=3$. However, $|W\cap X|=2$, so \spsc is violated.\medskip
	
	\textbf{SCR fails \spsc.} 	
	Consider the following profile with $n=9$ voters and $m=4$ candidates for $k=3$. 
	\begin{align*}
		&4:  a \succ d \succ b \succ c\\
		&3:  c \succ d \succ b \succ a\\
		&2:  b \succ c \succ d \succ a
	\end{align*}
	We first compute the SCR rule. In the first iteration of the first round, the solid coalition $S$ and corresponding candidate set $X$ that maximize the underrepresentation value are the five voters who prefer $\{b,c,d\}$ to $a$, achieving $\rho(W,S,X)=5$. In the next iteration, SCR narrows this down to the set $\{c\}$ which is top-ranked by $3$ voters, which constitutes the largest solid coalition that supports a strict subset of $\{b,c,d\}$. Therefore, candidate $c$ is the first selected, i.e., $W=\{c\}$.
	For the second round, we observe that the first $4$ voters form the solid coalition $S$ supporting $X=\{a\}$, which achieves the maximum underrepresentation value of $\rho(W,S,X)=4$. In particular, every larger solid coalition that includes $c$ has only an underrepresentation value of half its size. As a consequence, candidate $a$ is added to $W$, i.e., $W=\{a,c\}$. 
	In the third round, we observe that the last $5$ voters form the solid coalition $S$ supporting $X=\{b,c,d\}$, which achieves $\rho(W,S,X)=\frac{5}{1+1}=2.5$, which is the largest. Next, the last $2$ voters form a solid coalition $S$ supporting $X=\{b\}$ with $\rho(W,S,X)=2$. This value is maximal, so $b$ is the last selected candidate. Hence, SCR chooses the committee $W=\{a,b,c\}$ for our profile.
	
	Now consider the set of voters $S$ consisting of all the voters who strictly prefer $d$ to $b$, and let $\ell=3$, $X=\{d\}$ and $Y=\{b\}$. Note that $|S|=7 >3\cdot \frac{9}{3+1}= \ell \frac{n}{k+1}$. Since $d\succ_S b$, \spsc demands that $|W\cap X|\geq \min\{|X|,\ell-(m-|X|-|Y|)\}=\min\{1, 3-(4-1-1)\}=1$. However, we have $|W\cap X|=|\{a,b,c\}\cap \{d\}|=0$, which implies that the committee by the SCR rule fails \spsc.\medskip
	
	\textbf{STV satisfies \spsc.}
	We prove this claim by contradiction. To this end, fix a profile $R$ and a committee size $k$ and assume that the committee $W$ selected by STV fails \spsc. Thus, there exist a group of voters $S\subseteq N$, an integer $\ell$, and two disjoint candidate sets $X,Y\subseteq C$ such that
	$|S| > \ell\frac{n}{k+1}$, $X\succ_S Y$, and $|W\cap X|<\min\{|X|,\ell-(m-|X|-|Y|)\}.$
	We further let $Z=C\setminus (X\cup Y)$. Since $X\succ_S Y$ implies that $X\cap Y=\emptyset$, we have $|Z|=m-|X|-|Y|$. 
	Next, we define $t=\min\{|X|,\ell-(m-|X|-|Y|)\}=\min\{|X|,\ell-|Z|\}$. This means that $t\leq \ell-|Z|$, so $|S|>\frac{n}{k+1}(t+|Z|)$. Since $W$ violates \spsc, it holds that
	$|W\cap X|\leq t-1$, i.e., at most $t-1<|X|$ candidates from $X$ are selected in $W$. 
	
	Now, we consider the execution of STV on $R$. First, we note that STV must delete all candidates in $X$ from the profile, either to add them to the committee or to eliminate them as the candidates that are top-ranked least often. If this was not the case, STV would select $k$ candidates before exhausting the candidates in $X$. This means that the voters in $S$ never contribute to the selection of a candidate in $Y$, so we only ever delete voters from $S$ if a candidate from $X$ or $Z$ is selected. Since there are at most $(t-1)+|Z|$ such candidates in $W$, we delete at most $\frac{n}{k+1}\cdot (t-1+|Z|)$ voters from $S$. Hence, after the execution of STV, strictly more than $\frac{n}{k+1}$ voters in $S$ are left. However, by the definition of STV, the total weight of the remaining voters after its execution is $\frac{n}{k+1}$, a contradiction. This proves that it must be the case that each candidate in $X$ is deleted from the profile during the execution of STV.
	
	Further, since at most $t-1<|X|$ candidates of $X$ are selected, at least one candidate from $X$ must have been removed during the execution of STV while being top-ranked by the fewest voters. Consider the round where the last candidate from $X$ is eliminated without being added to the committee. In particular, this means that, in this round, no candidate is top-ranked by more than $\frac{n}{k+1}$ voters. Let $s_1$ denote the number of candidates in $X\cup Z$ that have been added to the committee in previous rounds of STV and let $s_2$ denote the number of candidates in $X\cup Z$ that have not yet been deleted. It holds that $s_1+s_2\leq t+|Z|$ because, after eliminating a candidate in $X$ in this round, every remaining candidate in $X$ will be added to the committee. Further, since a candidate from $X$ is still present in the considered round, the voters in $S$ only contributed to the selection of the candidates in $X\cup Z$ so far. Hence, the size of $S$ in the considered round satisfies $|S|>\frac{n}{k+1}(t+|Z|-s_1)\geq \frac{n}{k+1}s_2$. However, by the pigeonhole principle, this means that there must be a candidate in $X\cup Z$ that is top-ranked by more than $\frac{n}{k+1}$ voters, so STV will not delete a candidate without adding it to the committee in this round. This contradicts the choice of the considered round, so we now derive that our initial assumption that STV fails \spsc must have been wrong. 
\end{proof}

\subsection{Omitted Proofs from Section 4.3}\label{app:characterizations}

We will now prove \Cref{thm:simultaneousVeto}. To this end, we will show two auxiliary lemmas. First, we prove that, under a group-balanced veto sequence that contains every voter once (and thus is a permutation), each voter can be placed at the last position. 

\begin{lemma}\label{lem:group_balanced_permutation}
	Let $R$ be a profile with $n$ voters, and let $\succ^*$ be a preference relation reported by at least one voter in $R$. There exists a permutation $V=(v_1,\dots,v_n)$ of the voters such that voter $v_n$ reports $\succ^*$ and $|\{\ell'\in\{1,\dots,\ell\}\colon {\succ_{v_{\ell'}}}={\succ}\}|\leq\lceil \ell \frac{n(\succ)}{n}\rceil$ for every preference relation $\succ$ and every $\ell\in\{1,\dots,n\}$.
\end{lemma}
\begin{proof}
	Fix a profile $R$ and a preference relation $\succ^*$ as in the lemma. We construct the permutation $V=(v_1,\dots,v_n)$ using a simple greedy procedure. First, choose a voter $i^*$ reporting $\succ^*$ and reserve this voter for the final position $v_n$. We then fill the remaining positions sequentially. At each step, we choose an unassigned voter other than $i^*$ whose addition does not violate group-balancedness.
	To see that such a voter always exists, consider a position $\ell\in\{1,\dots,n-1\}$. For every preference relation $\succ$, we may use up to $\lceil \ell n(\succ)/n\rceil$ voters reporting $\succ$ among the first $\ell$ positions. Inductively, assume that among the positions $v_1,\dots,v_{\ell-1}$, every preference relation $\succ$ has been used at most $\lceil(\ell-1)n(\succ)/n\rceil$ times. This hypothesis holds trivially for $\ell=1$.
	Summing the bounds for position $\ell$ over all preference relations gives $\sum_{\succ}\lceil \ell n(\succ)/n\rceil\geq\ell$. Since only $\ell-1$ positions have been filled, at least one preference relation can occur once more without violating group-balancedness. If $\lceil \ell n(\succ^*)/n\rceil<n(\succ^*)$, then even if $\succ^*$ is the only one that can occur again, there is an unassigned voter reporting this preference relation other than $i^*$.
	It remains to consider the case where $\lceil \ell n(\succ^*)/n\rceil=n(\succ^*)$. Since $\ell<n$, we have $\ell n(\succ^*)/n<n(\succ^*)$, so the ceiling is strictly larger than the corresponding fraction. Consequently, $\sum_{\succ}\lceil \ell n(\succ)/n\rceil\geq\ell+1$. Thus, the total remaining capacity permitted by group-balancedness is at least two, whereas the reserved voter $i^*$ accounts for only one unfilled position. Hence, there is another unassigned voter who can be added without violating group-balancedness.
	The greedy construction can therefore be continued until positions $v_1,\dots,v_{n-1}$ have been filled. Placing $i^*$ in the final position then yields the required permutation.
\end{proof}

As our second auxiliary lemma, we will show that the set $W_{\mathrm{SV}}(R)$ of candidates that can be selected by Simultaneous Veto coincides with those that can be selected by very long group-balanced veto sequences.

\begin{lemma}\label{lem:simultaneous_veto_coupling}
	Consider a profile $R$ with $n$ voters and $m$ candidates, and let $s$ be a positive integer divisible by $(n!)^m$. Every group-balanced veto sequence $V=(v_1,\dots, v_{nms})$ selects a candidate in $W_{\mathrm{SV}}(R)$. Conversely, every candidate in $W_{\mathrm{SV}}(R)$ can be selected by such a sequence.
\end{lemma}

\begin{proof}
	We compare Serial Veto with a scaled execution of Simultaneous Veto in which every candidate has capacity $ns$ and every voter consumes at unit speed. Scaling the capacities in this way does not change the order in which candidates are exhausted or the set $W_{\mathrm{SV}}(R)$. Since the total capacity is $mns$ and the voters jointly consume at rate $n$, the process ends at time $ms$.
	We first observe that every candidate is exhausted at an integer time. Let $0=\tau_0<\tau_1<\dots<\tau_L=ms$ be the distinct exhaustion times; since at least one candidate is exhausted at every such time, $L\leq m$. For every candidate $c$, let $r_c(\tau_j)$ denote its remaining capacity immediately after the eliminations at time $\tau_j$, where the remaining capacity of an exhausted candidate is set to zero. We claim that $r_c(\tau_j)$ is divisible by $(n!)^{m-j}$.
	
	The claim holds at $j=0$, since $r_c(0)=ns$ and $s$ is divisible by $(n!)^m$. Suppose it holds at some $j<L$. Let $a_c(\tau_j)$ be the number of voters consuming candidate $c$ immediately after time $\tau_j$. Until the next exhaustion, these consumption rates remain constant. Hence, $\delta_j=\tau_{j+1}-\tau_j$ is the minimum of $r_c(\tau_j)/a_c(\tau_j)$ over candidates with $a_c(\tau_j)>0$. By the induction hypothesis, $r_c(\tau_j)$ is divisible by $(n!)^{m-j}$. Moreover, $a_c(\tau_j)\in\{1,\dots,n\}$ and therefore divides $n!$. It follows that every ratio $r_c(\tau_j)/a_c(\tau_j)$, and hence also their minimum $\delta_j$, is divisible by $(n!)^{m-j-1}$. Consequently, $r_c(\tau_{j+1})=r_c(\tau_j)-a_c(\tau_j)\delta_j$ is divisible by $(n!)^{m-j-1}$. This proves the claim by induction. In particular, every $\delta_j$, and thus every exhaustion time $\tau_j$, is an integer.
	
	Now, consider an arbitrary group-balanced Serial Veto sequence $V=(v_1,\dots, v_{nms})$. Since voters with the same preference relations are interchangeable for veto sequences, $V$ can be written as $V=(\pi_1,\dots,\pi_{ms})$, where every $\pi_h$ is a permutation of the voters. To see this, note that for every $\ell\in \{1,\dots, ms\}$ and preference relation $\succ$, it holds that $\lceil n\cdot \ell \cdot \frac{n(\succ)}{n}\rceil=\ell n(\succ)$. Consequently, group-balancedness requires that voters reporting $\succ$ appear exactly $\ell n(\succ)$ times in the prefix $(v_1,\dots, v_{n\ell})$. Moreover, since this claim also holds for $n(\ell+1)$, for every preference relation $\succ$ precisely $n(\succ)$ voters who report $\succ$ must appear in our veto sequence between the points $n\ell$ and $n(\ell+1)$. Since the choice of these voters does not matter, this implies the desired representation. Next, let $z_c(h)$ denote the remaining capacity of candidate $c$ after the first $h$ permutation blocks of the Serial Veto execution. We show by induction on $h$ that $z_c(h)=r_c(h)$ for every candidate $c$.
	
	The equality is immediate for $h=0$. Suppose that our claim holds after $h$ blocks, i.e., $z_c(h)=r_c(h)$. The two processes then have the same remaining candidates. Next, for every voter $i$, let $b_i$ denote her bottom-ranked candidate at the beginning of the permutation block $h+1$ and let $a_c(h)$ be the number of voters who bottom-rank $c$ at this point in time. Since all exhaustion times are integers, no candidate is exhausted strictly between $h$ and $h+1$ by Simultaneous Veto. It follows that $r_c(h)\geq a_c(h)$ for every remaining candidate $c$.
	We claim that every voter $i$ vetoes $b_i$ in the next Serial Veto block $\pi_{h+1}$. Suppose otherwise, and let $i$ be the first voter in this block who does not veto $b_i$. Every preceding voter has vetoed the candidate that was their least-preferred candidate at the beginning of the block. Thus, before voter $i$ acts, candidate $b_i$ has received at most $a_{b_i}(h)-1$ vetoes. Its remaining capacity is therefore at least $r_{b_i}(h)-(a_{b_i}(h)-1)\geq 1$, so $b_i$ is still available. Because candidates can only be removed during the block, $b_i$ remains voter $i$'s least-preferred available candidate, a contradiction.
	Hence, the block casts exactly $a_c(h)$ vetoes against each candidate $c$. Therefore, $z_c(h+1)=z_c(h)-a_c(h)=r_c(h)-a_c(h)=r_c(h+1)$, completing the induction. The Serial Veto and Simultaneous Veto executions thus have identical remaining capacities after every complete block, independently of the permutations used within the blocks.
	
	It remains to identify the winner. Every candidate outside $W_{\mathrm{SV}}(R)$ is exhausted at an integer time strictly before $ms$, and therefore no later than $ms-1$. By contrast, the candidates in $W_{\mathrm{SV}}(R)$ remain available until time $ms$. Hence, after the first $ms-1$ Serial Veto blocks, the remaining candidates are precisely those in $W_{\mathrm{SV}}(R)$. The last candidate exhausted in the final block must consequently belong to $W_{\mathrm{SV}}(R)$.
	Conversely, fix $w\in W_{\mathrm{SV}}(R)$. At time $ms-1$, candidate $w$ is consumed by at least one voter; choose one such voter $i^*$ and let ${\succ^*}={\succ_{i^*}}$. For each of the first $ms-1$ blocks, choose a permutation satisfying the prefix bounds from \Cref{lem:group_balanced_permutation}. For the final block, apply the same lemma with $\succ^*$ as the designated preference relation, and assign $i^*$ to its final position. The resulting sequence is group-balanced. Indeed, a prefix of length $hn+r$, where $0\leq r<n$, contains exactly $hn(\succ)$ voters reporting $\succ$ in its first $h$ complete blocks and at most $\lceil r n(\succ)/n\rceil$ such voters in the remaining positions. Their total is therefore at most $hn(\succ)+\lceil r n(\succ)/n\rceil=\lceil (hn+r)n(\succ)/n\rceil$.
	By our previous argument, the final block starts in the same state as the last unit interval of Simultaneous Veto, and every voter vetoes the candidate they consume during that interval. Thus, candidate $w$ has exactly as much remaining capacity as there are voters consuming it. All these voters except $i^*$ act before the final position, leaving $w$ with one unit of capacity. The final veto by $i^*$ therefore exhausts $w$, making it the Serial Veto winner.
\end{proof}

Based on our two lemmas, we are now ready to prove our characterization.

\simultaneousVeto*
\begin{proof}
	Let $f$ denote a homogeneous Serial Veto rule and consider an arbitrary profile $R$. First, suppose that $f(R)\subseteq W_{\mathrm{SV}}(R)$. We need to show that there is a group-balanced veto sequence that implements $f(R)$. To this end, we can directly use \Cref{lem:simultaneous_veto_coupling}: since $f(R)\subseteq W_{\mathrm{SV}}(R)$, this result guarantees that there is a group-balanced sequence of length $nms$ that selects $f(R)$, where we let $s=(n!)^m$.
	
	For the converse, suppose that $f$ is additionally group-balanced. In this case, let $sR$ denote a profile that consists of $s$ copies of $R$. By homogeneity, $f(sR)=f(R)$. Since $f$ is group-balanced, there is a group-balanced veto sequence $V=(v_1,\dots, v_{nmst})$ implementing $f(sR)$, where $t\in\mathbb{N}$ is some multiplier. Next, we note that relative share of voters reporting a given preference relation $\succ$ is the same in $R$ and $sR$. Thus, for every preference relation $\succ$ and every positive integer $h$, there appear at most $\lceil h\frac{s n(\succ)}{sn}\rceil=\lceil h\frac{n(\succ)}{n}\rceil$ voters reporting $\succ$ in the prefix $(v_1,\dots, v_h)$ of $V$. This means that $V$ can also be interpreted as a valid group-balanced veto sequence for the original profile $R$ by mapping the clones of each voter back to the original voter. Further, since the length of this sequence (relative to $R$) is $nmst$, we can again apply \Cref{lem:simultaneous_veto_coupling} to derive that the last vetoed candidate of this sequence is in $W_{\mathrm{SV}}(R)$. This shows that $f(R)\subseteq W_{\mathrm{SV}}(R)$.
\end{proof}

\section{Omitted Proofs from Section 5}\label{app:verification}

\violation*
\begin{proof}
	We first note that, by our duality framework, both statements of this proposition are equivalent to each other. Indeed, a committee $W$ fails \spsc for a profile $R$ and committee size $k$ if and only if $C\setminus W$ fails \pmvc for the reversed profile $\rev(R)$ and committee size $k'=m-k$. Furthermore, if there are two non-empty and disjoint candidate sets $X, Y\subseteq C$ and a voter set $S$ such that $X\succ_S Y$, $X\cap W=\emptyset$, and $|S|>(1+|C\setminus (X\cup Y)|)\cdot \frac{n}{k+1}$, then it holds that $X\subseteq C\setminus W$, $Y\mathrel{\rev(\succ_S}) X$, and $|S|>(1+|C\setminus (X\cup Y)|)\cdot \frac{n}{m-(m-k)+1}$. Thus, the size-$k$ committee $W$ meets the conditions of Statement (1) only if the size-$(m-k)$ committee $C\setminus W$ meets the conditions of Statement (2) for $\rev(R)$, and the converse direction follows analogously. Finally, by quantifying over all committee sizes, it follows that both statements are indeed logically equivalent to each other. 
	
	We will hence only prove the first claim. Thus, fix a size-$k$ committee $W$ and a profile $R$. First, if there are two non-empty and disjoint sets $X,Y\subseteq C$ such that $|\{i\in N\colon X\succ_i Y\}|>(1+|C\setminus(X\cup Y)|)\frac{n}{k+1}$ and $X\cap W=\emptyset$, we immediately have that $W$ fails \spsc. Indeed, this axiom demands that at least one candidate from $X$ is chosen, but $X\cap W=\emptyset$. 
	
	Conversely,  suppose that $W$ violates \spsc. This means that there exist a group of voters $S_0$ with $|S_0| > \ell_0 \frac{n}{k+1}$ and two disjoint candidate sets $X_0$ and $Y_0$ such that $X_0 \succ_{S_0} Y_0$ and $|W\cap X_0| < \min\{|X_0|, \ell_0 - |Z_0|\}$, where $Z_0=C\setminus (X_0\cup Y_0)$. Now let $X=X_0\setminus W$, $Y=Y_0$, $Z=C\setminus (X\cup Y)$, and $S=\{i\in N\colon X\succ_i Y\}$. We claim that these sets satisfy all conditions of our proposition. For this, we first note that $X$ is non-empty since $|W\cap X_0| < |X_0|$ implies that there is a candidate in $X_0$ that is not in $W$. Moreover, $X$ and $Y$ are disjoint because $X_0$ and $Y_0$ satisfy this condition. Third, $W\cap X=\emptyset$ by definition. Finally, since $|W\cap X_0| < \ell_0 - |Z_0|$, we have $\ell_0 > |Z_0|+|W\cap X_0|$. As all quantities involved are integers, it holds that $\ell_0\geq |Z_0|+|W\cap X_0|+1$. Moreover, since $X=X_0\setminus W$ and $Y=Y_0$, it follows that $Z=C\setminus((X_0\setminus W)\cup Y_0)=Z_0\cup (W\cap X_0)$. The two sets $Z_0$ and $W\cap X_0$ are disjoint, so $|Z|=|Z_0|+|W\cap X_0|$. Therefore, $\ell_0\geq |Z|+1$. Furthermore, it holds that $S_0\subseteq S$ since every voter in $S_0$ prefers every candidate in $X$ to every candidate in $Y$. In combination, this shows that $|S|\geq |S_0|> \ell_0\frac{n}{k+1}\geq (|Z|+1)\frac{n}{k+1}$. Hence, our sets satisfy all conditions listed in Statement (1).
\end{proof}

\verification*
\begin{proof}
	We will prove only the claim for \spsc but note that both are equivalent to each other. Indeed, a size-$k$ committee $W$ satisfies \spsc for a profile $R$ if and only if $C\setminus W$ satisfies \pmvc for the reverse profile $\rev(R)$. Hence, we can efficiently verify membership in the proportional multiwinner veto core if and only if we can efficiently verify whether a committee satisfies \spsc.
	
	Now, for the membership of \spsc verification in coNP, we note that, for no-instances, we can guess the witness for the \spsc violation under the profile and verify it. In particular, let $W$ denote a committee that fails \spsc under profile $R$. Then, a set of voters $S$, an integer $\ell$, and two non-empty subsets of candidates $X$ and $Y$ such that $|W\cap X| < \min(|X|, \ell-|C\setminus (X\cup Y)|)$, $|S| > \ell \frac{n}{|W|+1}$, and $X\succ_S Y$ constitute a polynomial-time verifiable certificate for the violation of \spsc. 
	
	To show the coNP-hardness, we reduce from the complement of the vertex cover problem on cubic graphs. The vertex cover problem is a classical NP-complete problem, and \citet{AlKa00a} showed that it remains hard even for cubic graphs; in fact, the corresponding minimum vertex cover problem is APX-complete. In the vertex cover problem on cubic graphs, we are given an undirected $3$-regular graph $G=(V,E)$ and an integer $\ell<|V|$, and are asked whether there exists a set of vertices $U\subseteq V$ such that $|U|\leq \ell$ and $e\cap U\neq\emptyset$ for every edge $e\in E$. The complement problem asks whether no such vertex cover exists. Given an instance $(G,\ell)$, we construct a preference profile $R$ and a committee $W$ such that $W$ fails \spsc if and only if $G$ admits a vertex cover of size at most $\ell$. Equivalently, $W$ satisfies \spsc if and only if $(G,\ell)$ is a yes-instance of the complement problem. We begin by defining the candidate set. Let $t=|V|$. Since $G$ is $3$-regular, we have $|E|=\frac{3}{2}t$. Let $q=t+\ell + 1$ and $C$ denote the candidate set. We first add $(3t-1)q + t-1$ candidates $T$ into $C$. We then add two disjoint candidate sets $D_X$ and $D_Y$, each of which has $q$ candidates. We finally add two sets of vertex candidates $C_X=\{x_v:v\in V\}$ and $C_Y=\{y_v:v\in V\}$. That is, for each vertex $v\in V$, we create two candidates $x_v$ and $y_v$ corresponding to $v$. Hence, we have 
	$C=T\cup C_X \cup C_Y \cup D_X \cup D_Y$ and there are in total $m=(3t+1)q+3t-1$ candidates. Next, we set the committee size to $k=3tq+t-1$ and consider the committee $W=T\cup D_Y$. 
	
	We next formulate the voters and their profile. We consider two types of voters, which we refer to as edge voters and vertex voters, respectively. Specifically, for each edge $e=(v,w)\in E$, we define two different types of voters:
	\begin{align*}
		\succ_e^1&=T\succ C_X\setminus \{x_v\} \succ D_X \succ y_w \succ x_v \succ D_Y \succ C_Y\setminus \{y_w\} \\
		\succ_e^2&=T\succ C_X\setminus \{x_w\} \succ D^{rev}_X \succ y_v \succ x_w \succ D_Y \succ C_Y\setminus \{y_v\}. 
	\end{align*}
	Here we use $D^{rev}_X$ to represent the reversed order of $D_X$. For any other set, the order over the alternatives will not matter. 
	For each edge $e=(v,w)$, we add $q$ copies of both of these preference relations, that is, for each edge $e$, there are $2q$ edge voters with two different types. So there are $3tq$ edge voters in total. 
	
	Regarding the vertex voters, for each vertex $v\in V$, we create one vertex voter with the preference
	\begin{align*}
		{\succ_v}=T\succ C_X\setminus \{x_v\} \succ y_v \succ D_X \succ D_Y \succ x_v \succ C_Y\setminus \{y_v\}.
	\end{align*}
	Therefore, there are in total $n=3tq + t=k+1$ voters. 
	
	We next show that $W$ fails \spsc if and only if there is a vertex cover of size at most $\ell$. First, suppose that there exists a valid vertex cover $U$ with size $|U|\leq \ell$. In this case, we define three disjoint sets of candidates, $X_U=D_X\cup \{x_v: v\notin U\}$, $Y_U=D_Y\cup \{y_v:v\notin U\}$, and $Z_U=T\cup  \{x_v,y_v:v\in U\}$. It holds that $C=X_U \cup Y_U \cup Z_U$. Now, consider the set of voters $S_U$ that consists of all $3tq$ edge voters and the  vertex voters corresponding to $U$. In particular, $|S_U|=3tq + |U|$. We first show that for each voter $i\in S_U$, $X_U\succ_i Y_U$. To see this, we first consider any edge voter and let $e=(v,w)$ denote the associated edge. We note that $D_X\cup C_X\setminus \{x_v\}\succ_e^1 D_Y\cup C_Y$ and that  $x_v\succ_e^1 D_Y\cup C_Y\setminus \{y_w\}$. Hence, the only inversion of $\succ_e^1$ could be that $y_w\succ_e^1 x_v$. However, since $U$ is a vertex cover, it holds that $v\in U$ or $w\in U$ and thus $y_w\in Z_U$ or $x_v\in Z_U$. An analogous argument applies for $\succ_e^2$, so every edge voter satisfies that $X_U\succ Y_U$. Next, for we turn to the vertex voters corresponding to vertices $v\in U$. Since both $x_v$ and $y_v$ are in $Z_U$ by definition, we directly infer for these voters that $X_U \succ_v Y_U$. Lastly, we observe that $|S_U|=3tq+|U|$ while $|Z_U|=|T|+2|U|=(3t-1)q+t-1 + 2|U|$. Using that $q=t+\ell+1$, we have 
	\begin{align*}
		|S_U| &=3tq+|U| = |Z_U|+q-|U|-t+1=  |Z_U|+1 + \ell + 1 - |U|.
	\end{align*}
	Since $|U| \leq \ell$ and $n=k+1$, we derive that $|S_U| > (|Z_U|+1)\frac{n}{k+1}$. Moreover, we have $W\cap X_U=\emptyset$. By \Cref{lem:violation_witness_strong_psc}, it follows that $W$ violates \spsc. 
	
	Conversely, suppose $W$ violates \spsc. We next show that there exists a valid vertex cover in $G$. By \Cref{lem:violation_witness_strong_psc}, since $W$ violates \spsc, there exists a voter set $S$, non-empty candidate sets $X$ and $Y$ such that $X\succ_S Y$, $W\cap X=\emptyset$, and $|S| > (|C\setminus (X\cup Y)|+1) \frac{n}{k+1}$. Without loss of generality, we assume that $S$ includes all the voters who prefer $X$ over $Y$ as any larger coalition also witnesses the violation. Denote by $Z=C\setminus (X\cup Y)$, and $U=\{v\in V: x_v \in Z \text{ or } y_v \in Z\}$. We next show that $U$ is a vertex cover of size at most $\ell$ in $G$. To this end, we first prove that our construction forces that all the edge voters are included in $S$, that $X$ is a subset of $C_X\cup D_X$, and that $Y$ is a subset of $C_Y\cup D_Y$. 
	\begin{claim}
		$S$ includes all the edge voters. 
	\end{claim}
	Suppose for contradiction that there is an edge voter excluded from $S$. Then, there are $q$ such voters with the same preference relation, all of which have to be excluded by the inclusion maximality of $S$. So we can upper-bound the size of $S$ by $|S| \leq n - q = (3t-1)q+t$. On the other hand, we recall that $|S| > (|Z|+1) \frac{n}{k+1}=|Z|+1$. We next lower-bound the size of $|Z|$ by showing that $T\subseteq Z$. To see this, we observe that $T\cap X=\emptyset$ as $W\cap X=\emptyset$ and $T \subseteq W$. Furthermore, it must be that $Y\cap T=\emptyset$ because every voter top-ranks the candidates in $T$. In more detail, if there was a candidate $y\in Y\cap T$, there is no set $X$ with $X\cap W=\emptyset$ such that $X\succ_i Y$ for any voter $i\in S$. So $T\cap Y=\emptyset$, and $T\subseteq Z$. Note that $|T|=(3t-1)q+t-1$ and $|Z|\geq |T|$. Hence $|S| > (3t-1)q+t$, contradicting that $|S| \leq (3t-1)q+t$. This implies all the edge voters must be included in $S$.
	
	\begin{claim}
		$X\subseteq C_X\cup D_X$ and $Y\subseteq C_Y\cup D_Y$. 
	\end{claim}
	
	To prove this claim, we first observe that $D_Y\cap X=\emptyset$ and $T\cap X=\emptyset$ since $W=D_Y\cup T$ and $W\cap X=\emptyset$. For the inclusion on $X$, it thus remains to show $C_Y\cap X=\emptyset$. Suppose not, so there exists some $y_v\in X\cap C_Y$. Since $G$ is a simple $3$-regular graph, there exists an edge $e=(u,w)$ that is not incident to $v$. For the corresponding edge voter with preference relation $\succ_e^1$,  we have $d\succ_e^1 y_v$ for every $d\in D_Y$. Since all the edge voters are in $S$ and $y_v\in X$, there exists no candidate $d\in D_Y\cap Y$; otherwise $d$ will be ranked before $y_v$. So $D_Y\cap Y=\emptyset$ and we further have $D_Y\subseteq Z$. Furthermore, by the argument in the previous claim, we have that $T\subseteq Z$ and therefore $|Z| \geq |T|+|D_Y|=k$. However, this contradicts that $|S| > |Z|+1$ as $|S|\leq n=k+1$. So, we conclude that $C_Y\cap X=\emptyset$ and finally $X\subseteq C_X\cup D_X$. 
	
	It remains to prove that $Y \subseteq C_Y \cup D_Y$. To this end, it suffices to show $Y$ is disjoint from $T,D_X$, and $C_X$. First, we have that $Y\cap T=\emptyset$ as reasoned before. Next, suppose for contradiction that $Y\cap D_X\neq \emptyset$ and let $d$ denote a candidate in $D_X\cap Y$. We first prove that if such a candidate $d$ exists, then $D_X\cap X = \emptyset$. If this was not the case, there exists a candidate $d'\in D_X \cap X$. Since, for every edge $e$, the edge voters reporting $\succ_e^1$ and $\succ_e^2$ rank the candidates in $D_X$ reversely, there exists an edge voter $i$ such that $d \succ_i d'$. Since all the edge voters are in $S$, $d\in Y$, and $d'\in X$, this contradicts $X\succ_S Y$. We thus conclude that $D_X\cap X = \emptyset$. Since we have already proven that $X\subseteq C_X\cup D_X$, we derive now that $X\subseteq C_X$. Observing $X\neq \emptyset$, consider any $x_v \in X$ and some edge $e \in E$ with $v \in e$. Then $D_X \succ x_v$ holds for either $\succ^1_e$ or $\succ^2_e$, so the corresponding edge voter $i$ satisfies $d \succ_i x_v$, which again contradicts $X \succ_S Y$. This means that $X \cap C_X = \emptyset$, so $X$ has to be empty. This contradicts the definition of $X$, so we derive that $Y\cap D_X=\emptyset$. Lastly, it remains to show that $Y\cap C_X=\emptyset$. Before proving this claim, we first show that $D_X\cap X\neq\emptyset$. Assume that $D_X\cap X=\emptyset$. Recall we have proven that $D_X\cap Y=\emptyset$ and $X\cup Y\cup Z=C$ and $X,Y,Z$ are disjoint. It follows that $D_X \subseteq Z$. We also have $T\subseteq Z$, which gives us a lower bound of size $|Z| \geq |T|+|D_X|=q + (3t-1)q+t-1=3tq+t-1$. On the other hand, by our assumption, $3tq+t=n \geq |S| > |Z|+1$ implies that $|Z| < 3tq+t-1$, a contradiction. Hence $D_X\cap X\neq \emptyset$ and we fix a candidate $d\in D_X\cap X$. Now suppose for contradiction that there exists a candidate $x_v\in Y\cap C_X$. Consider any edge $e=(u,w)$ with $v\not\in\{u,w\}$. It holds for the preference relation $\succ_e^1$ that $x_v \succ_e^1 D_X$, and so $x_v \succ_e^1 d$ as $d\in D_X$. However, since $x_v\in Y$ and $d\in X$, this again contradicts $X\succ_S Y$. Therefore, $C_X\cap Y=\emptyset$. This completes the proof that $X\subseteq D_X\cup C_X$ and $Y\subseteq D_Y\cup C_Y$. 
	
	After obtaining these two claims, we finally show that $U=\{v\in V: x_v \in Z \text{ or } y_v \in Z\}$ is a valid vertex cover for graph $G$ with $|U|\leq \ell$. Consider any edge $e=(v,w) \in E$ and the corresponding edge voters. There is an inverse pair of $y_w \succ_e^1 x_v$. We claim that either $x_v$ or $y_w$ is in $Z$. If this was not the case, then $x_v, y_w \in X\cup Y$. Recall that we have $X\subseteq D_X\cup C_X$ and $Y\subseteq D_Y\cup C_Y$. Because $x_v\in C_X$ and $Y \cap C_X=\emptyset$, $x_v$ must be in $X$. Similarly, we derive that $y_w$ must be in $Y$. However, all the edge voters are in $S$ and $X\succ_S Y$. This contradicts that $y_w\succ_e^1 x_v$ as $y_w\in Y$ while $x_v \in X$. Therefore, at least one of $x_v$ or $y_w$ is in $Z$. Consequently, by the definition of $U$, $v$ or $w$ is in $U$. This implies $U$ is a vertex cover of $G$. 
	
	It remains to show that $|U| \leq \ell$. 
	Note that each vertex $v$ corresponds to two candidates $x_v$ and $y_v$. We divide all the vertices into three disjoint subsets, $V_1=\{v: x_v, y_v \notin Z\}$, $V_2=\{v: |\{x_v,y_v\}\cap Z|=1\}$ and $V_3=\{v: x_v\in Z, y_v\in Z\}$ and observe that $|Z|\geq |T| + |V_2| + 2|V_3|$. On the other hand, we already know that all the edge voters are in $S$. Now consider the vertex voters in $S$. Notice that for each vertex voter in $S$, it must hold that $x_v\in Z$ and $y_v\in Z$; otherwise $X\succ_S Y$ fails by construction of $\succ_v$. In more detail, if $x_v\not\in Z$, it must be in $X$ since $Y\subseteq C_Y\cup D_Y$. Because the vertex voter corresponding to $v$ ranks all candidates in $D_Y$ ahead of $x_v$ and $X\subseteq C_X\cup D_X$, $D_Y$ must be a subset of $Z$. However, this means again that $|Z|\geq 3tq+t-1$, contradicting that $|S|>|Z|+1$. Similarly, if $y_v\not\in Z$, it must be in $Y$ as $X\subseteq C_X\cup D_X$. In turn, this means that $D_X\subseteq Z$ as the vertex voter prefers $y_v$ to all candidates in $D_X$ and $Y\subseteq C_Y\cup D_Y$. This again contradicts the size constraint on $Z$. Thus, we conclude that, if the vertex voter corresponding to $v$ is in $S$, then $x_v,y_v\in Z$. 
	Consequently, there are at most $|V_3|$ vertex voters in $S$ and $|S|\leq 3tq + |V_3|$. We also know that $|S| > |Z|+1$. Chaining the inequalities gives us $|T|+|V_2|+2|V_3| + 1 < 3tq+|V_3|$, or equivalently that $|V_2|+|V_3|<3tq-|T|-1$. By the definition of $U$, we have $U=V_2\cup V_3$ and $|U|=|V_2|+|V_3|$. So we derive that $|U| < 3tq-|T|-1=3tq- ((3t-1)q+t-1)-1=q-t=\ell + 1$ as $q=t+\ell+1$. Since $|U|$ and $\ell$ are integers, it follows that $|U|\leq \ell$, which completes the proof that $U$ is a valid vertex cover of $G$. 
\end{proof}

\fpt*
\begin{proof}
	We prove both claims of the proposition separately.\medskip
	
	\textbf{Claim (1):} To prove this claim, fix a size-$k$ committee $W$, a profile $R$, and a set of candidates $X$. If $X\cap W=\emptyset$, then $X$ cannot witness a violation of \pmvc, since $|X\setminus W|=|X|\geq \min(|X|,\ell)$ for every $\ell\in\mathbb{N}$. Hence, assume that $X\cap W\neq\emptyset$, so that $|X\setminus W|<|X|$. If $C\setminus X=\emptyset$, there is no non-empty set $Y$ with $Y\succ_S X$, so we may also assume that $C\setminus X\neq\emptyset$.
	We aim to find a voter set $S\subseteq N$ and a non-empty candidate set $Y\subseteq C\setminus X$ that maximize $\frac{m-k+1}{n}|S|+|Y|$ subject to $Y\succ_S X$. We claim that $W$ fails \pmvc for $X$ if and only if the maximum value satisfies $\frac{m-k+1}{n}|S|+|Y|>m-|X|+1+|X\setminus W|$. Indeed, this inequality is equivalent to $|S|>(1+|X\setminus W|+|C\setminus(X\cup Y)|)\frac{n}{m-k+1}$. Since $Y\succ_S X$ and $X\cap W\neq\emptyset$, the group $S$ can demand that $1+|X\setminus W|$ candidates from $X$ be unchosen, thereby witnessing a violation of \pmvc.
	Conversely, suppose that \pmvc is violated for $X$. Then there are non-empty sets $S'$ and $Y'$ and an integer $\ell$ such that $Y'\succ_{S'}X$, $|S'|>(\ell+|C\setminus(X\cup Y')|)\frac{n}{m-k+1}$, and $|X\setminus W|<\min(|X|,\ell)$. The latter inequality implies that $\ell\geq |X\setminus W|+1$. Consequently, $|S'|>(1+|X\setminus W|+|C\setminus(X\cup Y')|)\frac{n}{m-k+1}$, or, equivalently, $\frac{m-k+1}{n}|S'|+|Y'|>m-|X|+1+|X\setminus W|$. Thus, it suffices to maximize $\frac{m-k+1}{n}|S|+|Y|$ over sets $S\subseteq N$ and non-empty sets $Y\subseteq C\setminus X$ satisfying $Y\succ_S X$.
	
	To maximize our objective, for every candidate $y^*\in C\setminus X$, we construct a weighted directed graph $G_{y^*}=(V,E,c)$. Its vertices consist of a source $s$, a sink $t$, a candidate vertex $v_y$ for every $y\in C\setminus X$, and a voter vertex $u_i$ for every $i\in N$. There is an edge from $s$ to every candidate vertex $v_y$, with $c(s,v_y)=1$ if $y\neq y^*$ and $c(s,v_y)=2m$ if $y=y^*$. There is also an edge from every voter vertex $u_i$ to the sink $t$ with $c(u_i,t)=\frac{m-k+1}{n}$. Finally, for every $y\in C\setminus X$ and $i\in N$ such that voter $i$ does not prefer $y$ to every candidate in $X$, there is an edge from $v_y$ to $u_i$ with $c(v_y,u_i)=2m$.
	
	Let $(Z,V\setminus Z)$ be an $s$-$t$ minimum cut in $G_{y^*}$. We first observe that its value is strictly less than $2m$. To see this, consider the cut whose source side is $\{s,v_{y^*}\}\cup\{u_i\colon i\in N\}$ and whose sink side contains $t$ and all candidate vertices other than $v_{y^*}$. The edges crossing this cut have total cost $m-|X|-1+(m-k+1)=2m-|X|-k<2m$. Thus, no edge of capacity $2m$ crosses the minimum cut from $Z$ to $V\setminus Z$. Next, define $Y_{y^*}=\{y\in C\setminus X\colon v_y\in Z\}$ and $S_{y^*}=\{i\in N\colon u_i\in V\setminus Z\}$. Since the edge $(s,v_{y^*})$ has capacity $2m$, we have $y^*\in Y_{y^*}$, and therefore $Y_{y^*}\neq\emptyset$. Moreover, there is no edge $(v_y,u_i)$ with $y\in Y_{y^*}$ and $i\in S_{y^*}$, since such an edge would cross the cut from $Z$ to $V\setminus Z$ and would have capacity $2m$. It follows from the construction of the graph that every voter in $S_{y^*}$ prefers every candidate in $Y_{y^*}$ to every candidate in $X$. Hence, $Y_{y^*}\succ_{S_{y^*}}X$.
	
	The only edges crossing the minimum cut are unit-capacity edges $(s,v_y)$ for candidates $y\in(C\setminus X)\setminus Y_{y^*}$ and edges $(u_i,t)$ of capacity $\frac{m-k+1}{n}$ for voters $i\in N\setminus S_{y^*}$. Therefore, the value of the cut is $m-|X|-|Y_{y^*}|+\frac{m-k+1}{n}(n-|S_{y^*}|)=m-|X|+m-k+1-\left(|Y_{y^*}|+\frac{m-k+1}{n}|S_{y^*}|\right)$.
	Conversely, let $S'\subseteq N$ and $Y'\subseteq C\setminus X$ satisfy $y^*\in Y'$ and $Y'\succ_{S'}X$. Consider the cut with source side $Z'=\{s\}\cup\{v_y\colon y\in Y'\}\cup\{u_i\colon i\in N\setminus S'\}$ and sink side $V\setminus Z'=\{t\}\cup\{v_y\colon y\in(C\setminus X)\setminus Y'\}\cup\{u_i\colon i\in S'\}$. The edge $(s,v_{y^*})$ does not cross this cut because $y^*\in Y'$. Moreover, no incompatibility edge crosses from $Z'$ to $V\setminus Z'$, since $Y'\succ_{S'}X$. Thus, this cut has value $m-|X|-|Y'|+\frac{m-k+1}{n}(n-|S'|)=m-|X|+m-k+1-\left(|Y'|+\frac{m-k+1}{n}|S'|\right)$.
	By the optimality of the minimum cut, it follows that $|Y_{y^*}|+\frac{m-k+1}{n}|S_{y^*}|\geq |Y'|+\frac{m-k+1}{n}|S'|$. Hence, the minimum cut in $G_{y^*}$ identifies sets $S_{y^*}$ and $Y_{y^*}$ that maximize $\frac{m-k+1}{n}|S'|+|Y'|$ among all pairs satisfying $y^*\in Y'$ and $Y'\succ_{S'}X$.
	
	Finally, let $S$ and $Y$ maximize $\frac{m-k+1}{n}|S|+|Y|$ among all pairs $(S_{y^*},Y_{y^*})$ obtained from the graphs $G_{y^*}$ for $y^*\in C\setminus X$. We claim that $S$ and $Y$ maximize this expression among all pairs $S'\subseteq N$ and $\emptyset\neq Y'\subseteq C\setminus X$ satisfying $Y'\succ_{S'}X$. To see this, consider any such pair $(S',Y')$ and choose an arbitrary candidate $y^*\in Y'$. The algorithm considers the graph $G_{y^*}$, and the preceding argument gives $|Y_{y^*}|+\frac{m-k+1}{n}|S_{y^*}|\geq |Y'|+\frac{m-k+1}{n}|S'|$. Since $(S,Y)$ is chosen as the best pair returned by any of these minimum-cut computations, we have $|Y|+\frac{m-k+1}{n}|S|\geq |Y'|+\frac{m-k+1}{n}|S'|$. Therefore, $S$ and $Y$ are optimal. It follows that $W$ fails \pmvc for $X$ if and only if the sets returned by the algorithm satisfy $\frac{m-k+1}{n}|S|+|Y|>m-|X|+1+|X\setminus W|$.\medskip
	
	\textbf{Claim (2):} To prove this claim, fix a size-$k$ committee $W$, a profile $R$, and a set of voters $S$. We aim to find two non-empty sets $X\subseteq W$ and $Y\subseteq C$ that maximize $|X|+|Y|$ subject to $Y\succ_S X$. Given these sets, we claim that $W$ fails \pmvc for $S$ if and only if $|S|>(1+|C\setminus (X\cup Y)|)\frac{n}{m-k+1}$. First, suppose that $X$ and $Y$ satisfy this inequality. Then $S$, together with $X$ and $Y$, directly witnesses a violation of \pmvc, since $S$ is large enough to demand that at least one candidate from $X$ not be chosen. Conversely, suppose that $S$ witnesses that $W$ fails \pmvc. Then there are non-empty sets $X'$ and $Y'$ and an integer $\ell$ such that $Y'\succ_S X'$, $|S|>(\ell+|C\setminus (X'\cup Y')|)\frac{n}{m-k+1}$, and $|X'\setminus W|<\min(|X'|,\ell)$. Let $r=|X'\setminus W|$ and set $X''=X'\cap W$. Since $r<|X'|$, the set $X''$ is non-empty. Moreover, $Y'\succ_S X''$, and, since $r<\ell$, we have $\ell\geq r+1$. Consequently, $1+|C\setminus (X''\cup Y')|\leq \ell+|C\setminus (X'\cup Y')|$, and hence $|S|>(1+|C\setminus (X''\cup Y')|)\frac{n}{m-k+1}$. Thus, there are non-empty sets $X\subseteq W$ and $Y\subseteq C$ such that $Y\succ_S X$ and $|S|>(1+|C\setminus (X\cup Y)|)\frac{n}{m-k+1}$. Since $Y\succ_S X$ implies $X\cap Y=\emptyset$, the right-hand side of this inequality is decreasing in $|X|+|Y|$. Therefore, it suffices to identify non-empty sets $X\subseteq W$ and $Y\subseteq C$ that maximize $|X|+|Y|$ subject to $Y\succ_S X$.
	
	To find these sets, we first compute the set $P=\{(y,x)\in C^2\colon y\succ_S x\}$ of pairs of candidates such that all voters in $S$ prefer $y$ to $x$. We then iterate over all $x\in W$ and $y\in C$ such that $(y,x)\in P$. For each such pair $(y^*,x^*)$, we construct a weighted directed graph $G_{y^*x^*}=(V,E,c)$. Its vertices consist of a source $s$, a sink $t$, a committee vertex $v_x$ for every $x\in W$, and a candidate vertex $u_y$ for every $y\in C$. There is an edge from $s$ to every committee vertex $v_x$, with $c(s,v_x)=1$ if $x\neq x^*$ and $c(s,v_x)=2m$ if $x=x^*$. There is also an edge from every committee vertex $v_x$ to every candidate vertex $u_y$ such that $(y,x)\notin P$, with $c(v_x,u_y)=2m$. Finally, there is an edge from every candidate vertex $u_y$ to the sink $t$, with $c(u_y,t)=1$ if $y\neq y^*$ and $c(u_y,t)=2m$ if $y=y^*$.
	
	Next, we compute an $s$-$t$ minimum cut $(Z,V\setminus Z)$ in $G_{y^*x^*}$. We first observe that the cost of this cut is strictly less than $2m$. Indeed, consider the cut with source side $\{s,v_{x^*}\}\cup\{u_y\colon y\in C\setminus\{y^*\}\}$ and sink side $\{t,u_{y^*}\}\cup\{v_x\colon x\in W\setminus\{x^*\}\}$. The edges connected to $s$ and $t$ crossing this cut have total cost $(|W|-1)+(|C|-1)\leq 2(m-1)$. The only possible middle-layer edge crossing from the source side to the sink side would be $(v_{x^*},u_{y^*})$, but this edge does not exist because $(y^*,x^*)\in P$. Hence, this cut has cost at most $2(m-1)<2m$.
	It follows that no edge of capacity $2m$ crosses the minimum cut from $Z$ to $V\setminus Z$. Define $X_{y^*x^*}=\{x\in W\colon v_x\in Z\}$ and $Y_{y^*x^*}=\{y\in C\colon u_y\in V\setminus Z\}$. Since the edges $(s,v_{x^*})$ and $(u_{y^*},t)$ both have capacity $2m$, we have $x^*\in X_{y^*x^*}$ and $y^*\in Y_{y^*x^*}$. In particular, both sets are non-empty.
	Moreover, there can be no edge $(v_x,u_y)$ with $x\in X_{y^*x^*}$ and $y\in Y_{y^*x^*}$, since such an edge would cross from $Z$ to $V\setminus Z$ and would have capacity $2m$. By construction, this implies that $(y,x)\in P$ for every $x\in X_{y^*x^*}$ and $y\in Y_{y^*x^*}$. Therefore, $Y_{y^*x^*}\succ_S X_{y^*x^*}$. The only edges crossing the minimum cut are therefore unit-capacity edges of the form $(s,v_x)$ with $x\in W\setminus X_{y^*x^*}$ and $(u_y,t)$ with $y\in C\setminus Y_{y^*x^*}$. Thus, the cost of the cut is $|W|-|X_{y^*x^*}|+|C|-|Y_{y^*x^*}|$.
	
	Conversely, consider any sets $X'\subseteq W$ and $Y'\subseteq C$ such that $x^*\in X'$, $y^*\in Y'$, and $Y'\succ_S X'$. Define the source side $Z'=\{s\}\cup\{v_x\colon x\in X'\}\cup\{u_y\colon y\in C\setminus Y'\}$ and the sink side $V\setminus Z'=\{t\}\cup\{v_x\colon x\in W\setminus X'\}\cup\{u_y\colon y\in Y'\}$. No capacity-$2m$ edge crosses this cut from $Z'$ to $V\setminus Z'$: the two distinguished terminal edges do not cross because $x^*\in X'$ and $y^*\in Y'$, and no middle-layer edge $(v_x,u_y)$ with $x\in X'$ and $y\in Y'$ exists because $Y'\succ_S X'$. The edges $(v_x,u_y)$ with $x\in W\setminus X'$ and $y\in C\setminus Y'$ may cross the partition in the reverse direction, from $V\setminus Z'$ to $Z'$, but such edges do not contribute to the capacity of a directed cut. Hence, the cost of this cut is $|W\setminus X'|+|C\setminus Y'|=|W|+|C|-|X'|-|Y'|$. It follows that the minimum cut identifies sets $X_{y^*x^*}\subseteq W$ and $Y_{y^*x^*}\subseteq C$ that maximize $|X'|+|Y'|$ among all feasible pairs satisfying $x^*\in X'$, $y^*\in Y'$, and $Y'\succ_S X'$.
	
	Finally, let $X$ and $Y$ maximize $|X|+|Y|$ among all pairs $(X_{y^*x^*},Y_{y^*x^*})$ identified by this minimum-cut procedure for some $x^*\in W$ and $y^*\in C$ with $(y^*,x^*)\in P$. We claim that $X$ and $Y$ maximize $|X'|+|Y'|$ subject to $Y'\succ_S X'$, $\emptyset\neq X'\subseteq W$, and $\emptyset\neq Y'\subseteq C$. To see this, let $X'$ and $Y'$ be any maximizers of this objective, and choose $x'\in X'$ and $y'\in Y'$. Since $Y'\succ_S X'$, we have $(y',x')\in P$, so our algorithm computes a minimum cut in $G_{y'x'}$. By the preceding argument, the resulting sets satisfy $|X_{y'x'}|+|Y_{y'x'}|\geq |X'|+|Y'|$. Therefore, the selected sets $X$ and $Y$ satisfy $|X|+|Y|\geq |X'|+|Y'|$. Since every pair produced by the procedure is feasible, the reverse inequality follows from the optimality of $X'$ and $Y'$. Hence, $X$ and $Y$ are optimal, completing the proof.
\end{proof}

\section{Pareto Optimality}\label{app:Pareto}

As our last contribution, we turn to our claim on the (approximate) Pareto-optimality of the dual of sequential extensions of Serial Veto rules and STV in \Cref{rem:PO}. To formally study this property, for $\alpha\in [0,1]$, we say a candidate $x$ $\alpha$-Pareto-dominates another candidate $y$ in a profile $R$ if $\frac{|\{i\in N: x\succ_i y\}|}{n}\geq \alpha$, i.e., if at least an $\alpha$ fraction of the voters prefer $x$ to $y$. Moreover, a committee $W$ is called \emph{$\alpha$-Pareto optimal} in a profile $R$ if no candidate $x\in C\setminus W$ $\alpha$-Pareto-dominates a candidate $y\in W$. We note that, for $\alpha=1$, this is the common notion of (pairwise) Pareto optimality in committee voting. Further, $\alpha$-Pareto optimality gets stronger as $\alpha$ grows smaller, because it gets easier for candidates outside of $W$ to dominate those inside $W$. For example, for $\alpha=\frac{1}{2}$, $\alpha$-Pareto optimality requires that each candidate in the selected committee is preferred to every candidate outside of the committee by a strict majority of the voters. Moreover, in general, $(1-\alpha)$-Pareto optimality is impossible to attain for any $\alpha \geq \frac{1}{m}$, which follows by considering generalizations of the classical Condorcet cycle to $m$ alternatives.

Now, as claimed in \Cref{rem:PO}, we will next show that the committee chosen by STV satisfies $1$-Pareto optimality but fails this property for every $\alpha\leq 1-\frac{2}{2^{m-1}+1}$. Intuitively, this means that even for relatively small numbers of candidates, STV can elect candidates which are very close to being Pareto-dominated by unchosen candidates. 

\begin{proposition}
	\label{prop:STV_fails_eps_PO}
	The committee selected by STV always satisfies $1$-Pareto optimality. For every $m\geq 2$, there exists a profile with $m$ candidates, $2^{m-1}+1$ voters and $k=1$ such that STV fails $(1-\alpha)$-Pareto optimality for every $\alpha\geq \frac{2}{2^{m-1}+1}$.
\end{proposition}

\begin{proof}
	By~\citet{LLO+25a}, STV is proven to be lexicographically Pareto optimal, which immediately implies $1$-Pareto optimality in our context. We next show that STV can fail $(1-\alpha)$-Pareto optimality for every $\alpha\geq \frac{2}{2^{m-1}+1}$. To see this, we let $t=m-1$ and consider the following preference profile with $n=2^{t}+1$ voters and candidate set $C=\{x,y_1,\dots, y_t\}$ for the committee size $k=1$.
	\begin{align*} 
		2: & \, x, y_1, y_2, \dots, y_t \\ 
		1: & \,  y_1, x, y_2, \dots, y_t \\ 
		2: & \,  y_2, y_1, x, y_3, \dots, y_t \\ 
		4: & \, y_3,y_2,y_1, x, y_4, \dots, y_t\\
		& \, \vdots\\
		2^{t-1}: & \, y_t,y_{t-1}, \dots, y_1, x
	\end{align*}
	We claim that, in this profile, STV selects $x$. To see this, we note that, for $k=1$, STV only adds a candidate to the committee if it is top-ranked by more than $n/2=2^{t-1}+\frac{1}{2}$ voters. For $m \in \{2,3\}$, it is easy to verify that $x$ is selected (directly for $m=2$, and after deleting $y_1$ for $m=3$), so we assume $m \geq 4$ in the following. Hence, no candidate will be added to the committee initially and $y_1$ is deleted. The voter top-ranking $y_1$ then votes for $x$ and still no candidate can be added to the committee. Hence, $y_2$ will be deleted next, and the votes go again to $x$. This repeats until the candidates $y_1,\dots, y_{t-1}$ are deleted and $x$ is top-ranked by $2^{t-1}+1$ voters. Hence, $x$ wins in this round. However, we note that only two out of our $2^{t}+1$ voters prefer $x$ to $y_1$. This means that $y_1$ $(1-\alpha)$-Pareto-dominates $x$ for every $\alpha \geq \frac{2}{2^t+1}$, thus proving our claim.
\end{proof}

In contrast to STV, we next show that the committee chosen by the dual of the sequential extension of every Serial Veto rule satisfies $(1-\alpha)$-Pareto optimality for every $\alpha <\frac{1}{m}$. By the aforementioned cyclic construction, it is straightforward that this is the best one can hope for. Moreover, for committee election settings with few candidates and many voters, this guarantees that each candidate in the committee is preferred to every outside candidate by a non-negligible fraction of the voters.  

\begin{proposition}
	\label{prop:dual_seqSV_eps_PO}
	Let $f$ denote an arbitrary Serial Veto rule. The dual of the sequential extension of $f$ satisfies $(1-\alpha)$-Pareto optimality for every $\alpha<\frac{1}{m}$.
\end{proposition}
\begin{proof}
	Fix a Serial Veto rule $f$. Let $f^{\mathrm{seq}}$ denote its sequential extension, and $g$ denote the dual of $f^{\mathrm{seq}}$. Furthermore, consider an arbitrary profile $R$ and a committee size $k$, and let $W=g(R,k)$ denote the committee selected by $g$. Let $x$ denote an arbitrary candidate in $W$ and let $y\in C\setminus W$. By the definition of a dual rule, it holds that $f^{\mathrm{seq}}(\rev(R), m-k)=C\setminus W$, so $y\in f^{\mathrm{seq}}(\rev(R), m-k)$. Next, let $\ell$ denote the round such that $y\in f^{\mathrm{seq}}(\rev(R), \ell)$ and $y\not\in f^{\mathrm{seq}}(\rev(R), \ell-1)$. Further, let $\rev(R')=\rev(R)|_{C\setminus f^{\mathrm{seq}}(\rev(R), \ell-1)}$ denote the remaining profile when $y$ is chosen, i.e., it holds that $f(\rev(R'))=\{y\}$. We note that $x$ is present in $\rev(R')$ since $x\not\in f^{\mathrm{seq}}(\rev(R), m-k)$. 
	Finally, let $S$ denote the set of voters who prefer $x$ to $y$ in $\rev(R)$ and assume for contradiction that $|S|>n\frac{m-1}{m}$. Since $\rev(R')$ contains only $m' \leq m$ candidates, it follows that $|S| > n\frac{m-1}{m} \ge n\frac{m'-1}{m'}$. This means that $y$ is not in the proportional singlewinner veto core of $\rev(R')$ since $S$ is large enough to veto the $m'-1$ candidates other than $x$ and all of them prefer $x$ to $y$. However, this contradicts the insight that the candidate chosen by a Serial Veto rule always belongs to the proportional singlewinner veto core \citep{Moul82a,KiKe23a}. Hence, it must be that $|S|\leq n\frac{m-1}{m}$, so at least $\frac{n}{m}$ voters prefer $y$ to $x$ in $\rev(R)$. In turn, for $R$, this means that at least $\frac{n}{m}$ voters prefer $x$ to $y$. Thus, $y$ cannot $(1-\alpha)$-Pareto-dominate $x$ for any $\alpha<\frac{1}{m}$, which proves our claim. 
\end{proof}

\end{document}